\documentstyle[aps,eqsecnum,preprint,tighten,]{revtex}

\begin{document}

\title{Kerman-Klein-D\"{o}nau-Frauendorf model for odd-odd nuclei: formal theory}

\author{Abraham Klein
\footnote{Email:{aklein@dept.phys.upenn.edu}}  and 
Pavlos Protopapas
\footnote{Email:{pavlos@nscp.upenn.edu}}}
\address{Department of Physics and Astronomy, University of Pennsylvania,
Philadelphia, PA 19104-6306, USA}
\author{ 
Stanis{\l}aw G. Rohozi\'{n}ski
\footnote{Email:{Stanislaw-G.Rohozinski@fuw.edu.pl}}}
\address{Institute of Theoretical Physics, Warsaw University, Ho\.{z}a 69, PL-00-681 Warsaw, Poland}
\author{and Krzysztof Starosta
\footnote{Email:{Starosta@nuclear.physics.sunsysb.edu}}}
\address{Department of Physics and Astronomy, State University of 
New York at Stony Brook, Stony Brook, New York 11794-3800}

\maketitle

\bigskip
\begin{abstract} 

The Kerman-Klein-D\"{o}nau-Frauendorf (KKDF) model is a linearized version of the non-linear Kerman-Klein (equations of motion) formulation of the nuclear many-body problem.  In practice, it is a generalization of the standard core-particle coupling model
that, like the latter, provides a description of the spectroscopy of odd nuclei in terms of the corresponding properties of neighboring even nuclei
and of single-particle properties, that are the input parameters of the model.  A divers sample of recent applications attest to the usefulness of the model. In this paper, we first present a concise general review of the fundamental equations and properties
of the KKDF model.  We then derive a corresponding formalism for odd-odd nuclei with proton-neutron number $(Z, N)$ that relates their properties to those of the four neighboring even nuclei $(Z+1,N+1)$, $(Z-1,N+1)$, $(Z+1,N-1)$, and $(Z-1,N-1)$, all of
which are required if one is to include both multipole and pairing forces.  We treat these equations in two ways.  In the first, we make essential use of the solutions of the neighboring odd nucleus problem, as obtained by the KKDF method. In the second, we relate the properties of the odd-odd nucleus directly to those of the even nuclei.  For both choices, we derive equations of motion, normalization conditions, and an expression for transition amplitudes.  We also resolve the problem
of choosing the subspace of physical solutions that arises in an equations of motion approach that includes pairing interactions.

\end{abstract}
\bigskip


\newpage
 
\section{Introduction}

The Kerman-Klein-D\"{o}nau-Frauendorf (KKDF) model for odd nuclei was introduced
and applied \cite{DF1,DF2,DF3,POL1,DF4,DF5} as a semi-phenomenological approximation to the Kerman-Klein (KK) self-consistent formulation of the equation of motion approach to nuclear collective motion \cite{KK1,KK2,KK3,KK4,KK5}.  As such it generalizes phenomenological core-particle coupling models, to which it can be shown to reduce in various limits \cite{cpc}.
The past decade has witnessed further development of the theory and additional 
applications \cite{pav1,pav2,pav3,pav4,pav5,Lim,pav6,pav7,PRC} including, for example, a suggested solution of the Coriolis attenuation problem \cite{pav5,Lim}.   A review of this more recent work is in preparation \cite{KP}.

The main purpose of this paper is to show that a formalism of the KKDF type, at the same level of completeness as for odd nuclei can be constructed for odd-odd nuclei.  A first important step in this direction has already been made by Starosta {\it et al}
 who have applied a restricted version of the formalism to 
the phenomenon of chirality in odd-odd triaxial nuclei \cite{STAR}.  The restriction is the ommission of pairing interactions.  When the latter are included, we face,
among other difficulties, the problem that the manifold of solutions is four times the size of the manifold of physical solutions.  More recently Koike {\it et al} \cite{Kioke}, these
authors have applied an approximate form of the formalism developed in Sec.~III.

As a preliminary step, in Sec.~II, we review the KKDF program for odd nuclei.
We do this in a form which is both more general and more concise than can be found
in our previously published work, and which sets the stage for the work on odd-odd nuclei that follows.  It is more general in the sense that the equations are not restricted to deformed nuclei.  It is more concise in the sense that in our published work,
 we have described up to three different methods for choosing the
physical subspace of solutions, whereas here we choose that one of these methods that should work in all cases and is, in any event, the simplest to implement.

In Sec.~III, we present the first of two methods that can be used for odd-odd nuclei.  We refer to this as the sequential method in that it solves the problem by two successive applications of the KKDF approach to odd nuclei, utilizing the solutions for neighboring odd nuclei to derive equations for an odd-odd nucleus relative to it neighboring odd nuclei, so that the method involves only single-particle coefficients of fractional parentage (CFP).  In Sec.~IV, in an approach that treats the pair of odd particles symmetrically, we derive a set of eigenvalue equations and attendant orthonormalization conditions for 
two-particle (proton-neutron) coefficients of fractional parentage.  These amplitudes relate the given odd-odd nucleus to any of
four neighboring even nuclei.  For both approaches, we solve the problem of choosing the physical subspace of solutions.  Finally we derive for each case formulas for single-particle transition matrix elements that clearly separate collective and single-particle contributions.

\section{Review of model for odd nuclei}

\subsection{Equations of motion}

In this section we shall derive a version of the Kerman-Klein (KK) equations 
based on the Hamiltonian (\ref{kkp1.1}) given below.  
These equations, when taken literally, define a nonlinear problem for the
self-consistent study of the properties of an odd nucleus and of its
immediate even neighbors.   However, the version of the theory developed here, referred to as the Kerman-Klein-D\"{o}nau-Frauendorf (KKDF) model, has
a more modest goal.  This goal is achieved by making such
further approximations as to reduce the problem to a linear eigenvalue
problem for the properties of odd nuclei, assuming the required properties
of the neighboring even nuclei to be known.  This can be done only if 
the Hamiltonian can be chosen of sufficiently simple form that the matrix
elements of its ingredient multipole and pairing operators can be related
to observed properties of the even neighbors.
Even with such simplification, the resulting
theory generalizes previous core-particle coupling models.

We start with a shell-model Hamiltonian of the form
\begin{eqnarray}
H&=& \sum_\alpha h_a a_\alpha^{\dag} a_\alpha  
+\frac{1}{2}F_{\alpha\gamma\delta\beta}a_\alpha^{\dag}a_\gamma a_\beta^{\dag}
a_\delta +\frac{1}{2}G_{\alpha\beta\gamma\delta}a_\alpha^{\dag}a_\beta^{\dag}
a_\delta a_\gamma \nonumber \\
&=& \sum_\alpha h_a a_\alpha^{\dag} a_\alpha  
+ \frac{1}{2} \sum_{abcd}\sum_{LM_L}F_{acdb}(L)
B^{\dag}_{LM_L}(ac)B_{LM_L}(db) \nonumber \\ 
&& + \frac{1}{2} \sum_{abcd}\sum_{LM_L} G_{abcd}(L)
A^{\dag}_{LM_L}(ab)A_{LM_L}(cd).     \label{kkp1.1}
\end{eqnarray}
Here $h_a$ are the spherical single-particle energies referred to the
nearest closed shell, $\alpha$ refers to the standard set of single-particle
quantum numbers, including in particular the pair $(j_a ,m_a)$ and $a$
refers to the same set with $m_a$ omitted.
The charge conservation requirement means that only the matrix elements of interactions $F$ and $G$ which fulfill the condition
\begin{equation}
q_a + q_b=q_c+q_d,  \label{cc}
\end{equation}
where $q_a$ is the electric charge of a nucleon with the set of quantum numbers $a$, do not vanish and enter in
the Hamiltonian of Eq. (\ref{kkp1.1}). In the KKDF model we assume additionally two more restrictive conditions for the interaction matrix elements, namely: 
\begin{enumerate}
\item the charge exchange interactions are excluded, i.e. $q_a=q_c$ and $q_b=q_d$ for non vanishing matrix elements $F_{\alpha\gamma\delta\beta}$, 
\item only the pairs of like nucleons are correlated, i.e.  $q_a=q_b$ and $q_c= q_d$ for non vanishing  matrix elements $G_{\alpha\beta\gamma\delta}$.
\end{enumerate}
  $B^{\dag}_{LM_L}$ is the 
particle-hole multipole operator,
\begin{eqnarray}
B^{\dag}_{LM_L}(ab)&\equiv &\sum_{m_a m_b}s_\beta (j_a m_a j_b -m_b|LM_L)
a^{\dag}_\alpha a_\beta  \nonumber \\
&=& (-1)^{j_a +j_b -M_L+1}B_{L-M_L}(ba),    \label{kkp1.2}
\end{eqnarray}
and $A^{\dag}_{LM_L}$ is the particle-particle multipole operator,
\begin{equation}
A^{\dag}_{LM_L}(ab)\equiv \sum_{m_a m_b} (j_a m_a j_b m_b|LM_L)
a^{\dag}_\alpha a^{\dag}_{\beta},   \label{kkp1.3}
\end{equation}
where $(j_1 m_1 j_2 m_2|jm)$ is a Clebsch-Gordon (CG) coefficient,
$s_\alpha = (-1)^{j_a -m_a}$.  The coefficients $F$ are the particle-hole
matrix elements,
\begin{eqnarray}
&&F_{acdb}(L)\equiv \sum_{m's}s_\gamma s_\beta (j_a m_a j_c -m_c|LM_L)
\nonumber \\
&&\times(j_d m_d j_b -m_b|LM_L) F_{\alpha\gamma\delta\beta},  \label{kkp1.4}
\end{eqnarray}
and $G$ the particle-particle matrix elements
\begin{eqnarray}
&&G_{abcd}(L)\equiv \sum_{m's} (j_a m_a j_b m_b|LM_L)  \nonumber \\
&&\times(j_c m_c j_d m_d|LM_L)G_{\alpha\beta\gamma\delta}.
\label{kkp1.5}
\end{eqnarray}
Assuming the matrices $F$ and $G$ are real, we have 
\begin{eqnarray}
F_{acdb}(L)&=& F_{dbac}(L),  \label{kkp1.F} \\
G_{acdb}(L)&=& G_{dbac}(L)  \nonumber \\ 
&=& (-1)^{j_a +j_c -L +1}G_{cadb} \nonumber \\
&=& (-1)^{j_b +j_d -L +1}G_{acbd}.  \label{kkp1.G}
\end{eqnarray}

The task is to obtain equations for the states and energies of
an odd nucleus assuming that properties of immediately neighboring even nuclei
are known. The states of the odd nucleus (particle number A) are designated below
as $|J \mu\nu\rangle$, where $\nu$ denotes all quantum numbers besides the
angular momentum $J$ and its projection $\mu$. The states of the neighboring
even nuclei with particle numbers $(A\pm 1)$ are written, in a parallel
notation, as $|IMn(A\pm 1)\rangle$.  The corresponding eigenvalues are 
$E_{J\nu}$ and $E_{In}^{(A\pm 1)}$, respectively.  We first obtain the operator equations of motion (EOM), bar indicating reversal of the sign of the single-particle magnetic quantum number, 
\begin{eqnarray}
&&{}[a_{\bar{\alpha}},H]= h_a^{\prime} a_{\bar{\alpha}}   \nonumber \\
&&+  \sum_{bd\gamma}\sum_{LM}s_{\bar{\gamma}} (j_a -m_a j_c m_c|LM)
\bar{F}_{acdb}(L)a_{\bar{\gamma}} B_{LM}(db)  \nonumber \\
&&+  \sum_{bd\gamma}\sum_{LM} (j_a -m_a j_c m_c|LM)G_{acbd}(L)
a^{\dag}_{\gamma} A_{LM}(bd),  \label{ppk1.eom1} \\
&&{}[a^{\dag}_{\alpha},H]= -h_a^{\prime\prime} a^{\dag}_{\alpha} 
\nonumber \\
&&-  \sum_{bd\gamma}\sum_{LM}s_{\gamma} (j_a m_a j_c -m_c|LM)
a^{\dag}_{\gamma}\bar{F}_{acdb}(L) B^{\dag}_{LM}(db) \nonumber \\
&&  -\sum_{bd\gamma}\sum_{LM} (j_a m_a j_c -m_c|LM)a_{\bar{\gamma}}
G_{acbd}(L)A^{\dag}_{LM}(bd).  \label{ppk1.eom2} 
\end{eqnarray}
Here
\begin{eqnarray}
\bar{F}_{acdb}&=&\frac{1}{2}(F_{acdb}+(-1)^{j_a +j_{c}+j_b +j_{d}}
F_{bdca})=F_{acdb},   \label{barf}\\
h_a^{\prime}&=&h_a -\frac{1}{2}\sum_{Lj_c}F_{caca}(L)\frac{2L+1}{2j_a +1},
\label{ppk1.6}\\
h_a^{\prime\prime}&=&h_a +\sum_{Lj_c}\frac{2L+1}{2j_a +1}(2G_{acac}
+\frac{1}{2}F_{acac}).   \label{ppk1.7}
\end{eqnarray}
In consequence of (\ref{barf}), we may replace $\bar{F}$ by $F$.

The appearance of different single-particle energies in the two equations may be traced to the rearrangement of operators required to have the EOM in a form necessary to achieve our aims.  This requires, as we shall see below, that
the multipole and pairing operators occur on the extreme right.
The matrix elements of these equations provide expressions that determine the 
single-particle coefficients of fractional parentage (CFP),
\begin{eqnarray}
V_{J\mu\nu}(\alpha IMn)& =& \langle J\mu\nu|a_{\bar{\alpha}}|IMn(A+1)\rangle , 
\label{ppk1.vee} \\
U_{J\mu\nu}(\alpha IMn)&=& \langle J\mu\nu|a^{\dag}_{\alpha}|
IMn(A-1)\rangle.  \label{ppk1.you}
\end{eqnarray}
To find equations for these quantities, we form the necessary matrix
elements of the EOM and evaluate the interaction terms by inserting the 
completeness relation for the states of the appropriate even nuclei between the single-fermion operators and the multipole or pair operators.  

In terms of a convenient and physically
meaningful set of energy differences and sets of multipole fields
and pairing fields defined below, we thereby obtain generalized matrix
equations of the Hartree-Bogoliubov form   
\begin{eqnarray}
&&{\cal E}_{J\nu}V_{J\mu\nu}(\alpha IMn)  \nonumber\\
&&= (\epsilon^{\prime} +\omega^{(A+1)} 
+\Gamma^{(A+1)})_{{\bar\alpha} IMn,{\bar\gamma} I^{\prime} M' n'}
V_{J\mu\nu} (\gamma I' M' n')   \nonumber \\
&&+\Delta_{{\bar\alpha} IMn,\gamma I' M' n'}U_{J\mu\nu}(\gamma I'M' n'), 
\label{ppk1.hfb1} \\
&&{\cal E}_{J\nu}U_{J\mu\nu}(\alpha IMn)  \nonumber \\
&&= (-\epsilon^{\prime\prime} +\omega^{(A-1)} 
-\Gamma^{(A-1)\dag})_{\alpha IMn,\gamma I' 
M' n'}U_{J\mu\nu}(\gamma I' M' n')   \nonumber \\
&&-\Delta^{\dag}_{\alpha IMn,\bar{\gamma} I' M' n'}
V_{J\mu\nu}(\gamma I'M' n').   \label{ppk1.hfb2} 
\end{eqnarray}
Here
\begin{eqnarray}
&&{\cal E}_{J\nu} = -E_{J\nu} +\frac{1}{2}(E_0^{(A+1)}+E_0^{(A-1)}), 
\label{ppk1.def1}  \\
&&\epsilon^{\prime}_{\alpha IMn,\gamma I'M'n'} = \delta_{\alpha\gamma}
\delta_{II'}\delta_{MM'}\delta_{nn'}(h_a^{\prime} -\lambda_A), 
\label{ppk1.def2} \\
&&\lambda_A = \frac{1}{2}(E_0^{(A+1)}-E_0^{(A-1)}),  \label{ppk1.def3} \\
&&\omega^{(A\pm 1)}_{\alpha IMn,\gamma I'M'n'} = \delta_{\alpha\gamma}
\delta_{II'}\delta_{MM'}\delta_{nn'}(E_{In}^{(A\pm 1)}-E_0^{(A\pm 1)}),
\label{ppk1.def4} \\
&&\Gamma^{(A\pm 1)}_{\alpha IMn,\gamma I'M'n'} = \sum_{L}
\sum_{bd}s_\gamma (j_a m_a j_c -m_c|LM_L)  \nonumber \\
&&\times F_{acdb}(L)
 \langle I'M'n'(A\pm 1)|B_{LM_L}(db)|IMn(A\pm 1)\rangle \label{ppk1.def5}\\
&&\Delta_{\alpha IMn,\gamma I'M'n'} =\sum_{L}\sum_{bd}
(j_a m_a j_c m_c|LM_L) \nonumber \\
&& \times G_{acdb}(L) 
\langle I'M'n'(A- 1)|A_{LM_L}(db)|IMn(A+1)\rangle. \label{ppk1.def6}
\end{eqnarray}
Furthermore $E_0^{(A\pm 1)}$ refer to the ground state energies of the 
neighboring even nuclei, the matrix elements of $\Gamma^{\dag}$ 
are derived from those of (\ref{ppk1.def5})
simply by the replacement of the operator $B$ by $B^{\dag}$, and the matrix
elements of $\Delta^{\dag}$ are similarly derived from those of $\Delta$
by the replacement of $A$ by $A^{\dag}$ together with the interchange
$A\pm 1\rightarrow A\mp 1$.  Finally $\epsilon^{''}_a$ is obtained from
$\epsilon^{'}_a$ by the replacement of $h_a^{'}$ by $h_a^{''}$.

To specify fully solutions of the equations given above, we must develop
orthonormalization conditions for the CFP that fix their scale. Orthogonality
conditions can be derived from the equations of motion themselves.  A
normalization condition, on the other hand, is obtained by
taking a suitable matrix element of the summed anticommutator,
\begin{eqnarray}
\sum_\alpha \{a_\alpha , a^{\dag}_\alpha\} &=& \Omega, \\
\Omega = \sum_{j_a}(2j_a +1).
\end{eqnarray}
We thus find
\begin{equation}
\frac{1}{\Omega}\sum_{\alpha IMn} [|U_{J\mu\nu}(\alpha; IMn)|^2
+|V_{J\mu\nu}(\alpha; IMn)|^2] =1.     \label{kkp1.sum}
\end{equation}

\subsection{Equations for reduced matrix elements} 

To apply the Wigner-Eckart theorem to obtain the EOM for the reduced matrix elements, we utilize the following definitions for the latter (which suppress nucleon number):
\begin{eqnarray}
V_{J\mu\nu}(\alpha IMn)&=&(-1)^{j_a-m_a}  (IMj_am_a|J\mu)v_{J\nu}(aIn),  \label{kkdo1.1}\\
U_{J\mu\nu}(\alpha IMn)&=&(IMj_am_a|J\mu)u_{J\nu}(aIn), \label{kkdo1.2} \\
(I'M'n'|B_{LM_L}(bb')|IMn)&=&(-1)^{L-M_L}(IML-M_L|I'M')(I'n'||B_L(bb')||In),
\label{kkdo1.4a} \\
(I'M'n'|A_{LM_L}(bb')|IMn)&=&(-1)^{L-M_L}(IML-M_L|I'M')(I'n'||A_L(bb')||In),
\label{kkdo1.5} \\
(I'M'n'|B^{\dag}_{LM_L}(bb')|IMn)&=&(IMLM_L|I'M')(I'n'||B_L^{\dag}(bb')||In),
\label{kkdo1.3} \\
(I'M'n'|A^{\dag}_{LM_L}(bb')|IMn)&=&(IMLM_L|I'M')(I'n'||A_L^{\dag}(bb')||In).
\label{kkdo1.4} 
\end{eqnarray}
Assuming the reality of the multipole and pairing matrix elements, we also have
\begin{eqnarray}
(I'M'n'|B_{LM_L}|IMn)&=&(I'M'LM_L|IM)(In||B_L^{\dag}(bb')||I'n'),
\label{kkdo1.4aa} \\
(I'M'n'|A_{LM_L}|IMn)&=&(I'M'LM_L|IM)(In||A_L^{\dag}(bb')||I'n').
\label{kkdo1.5a} 
\end{eqnarray}

With the help of these definitions, we can transform Eqs.~(\ref{ppk1.hfb1}) and
(\ref{ppk1.hfb2}) into the forms
\begin{eqnarray}
{\cal E}_{J\nu}v_{J\nu}(aIn)&=&(\epsilon^{\prime}_a+\omega^{(A+1)}_n)v_{J\nu}
(aIn) \nonumber\\
&&+\sum_{a'I'n'}\Gamma_J^{(A+1)}(aIn|a'I'n')v_{j\nu}(a'I'n')  \nonumber\\
&&+\sum_{a'I'n'}\Delta_J(aIn|a'I'n')u_{J\nu}(a'I'n'),  \label{kkdo1.6}\\
{\cal E}_{J\nu}u_{J\nu}(aIn)&=&(-\epsilon^{\prime\prime}_a+\omega^{(A-1)}_n)u_{J\nu}(aIn) \nonumber\\
&&-\sum_{a'I'n'}\Gamma_J^{\dag (A-1)}(aIn|a'I'n')u_{j\nu}(a'I'n')  \nonumber\\
&&+\sum_{a'I'n'}\Delta_J^{\dag}(aIn|a'I'n')v_{J\nu}(a'I'n'),  \label{kkdo1.7}
\end{eqnarray}
where
\begin{eqnarray}
\Gamma_J^{(A+1)}(aIn|a'I'n')&=&\sum_{Lbb'}(-1)^{j_a +J+I}
\sqrt{(2I'+1)(2L+1)}\left\{\begin{array}{ccc}I&I'&L\\j_{a'}&j_a&J\end{array}
\right\} \nonumber \\
&&\times F_{aa'bb'}(L)(I'n'||B_L(bb')||In),  \label{kkdo1.8} \\
\Delta_J(aIn|a'I'n')&=&\sum_{Lbb'}(-1)^{j_a +J+I+1}
\sqrt{(2I'+1)(2L+1)}\left\{\begin{array}{ccc}I&I'&L\\j_{a'}&j_a&J\end{array}
\right\} \nonumber \\
&&\times G_{aa'bb'}(L)(In||A^{\dag}_L(bb')||I'n'),  \label{kkdo1.9} \\
\Gamma_J^{\dag(A-1)}(aIn|a'I'n')&=&\sum_{Lbb'}(-1)^{j_a +J+I}
\sqrt{(2I'+1)(2L+1)}\left\{\begin{array}{ccc}I&I'&L\\j_{a'}&j_a&J\end{array}
\right\} \nonumber \\
&&\times F_{aa'bb'}(L)(I'n'||B^{\dag}_L(bb')||In),  \label{kkdo1.10} \\
\Delta_J^{\dag}(aIn|a'I'n')&=&\sum_{Lbb'}(-1)^{j_a +J+I+1}
\sqrt{(2I'+1)(2L+1)}\left\{\begin{array}{ccc}I&I'&L\\j_{a'}&j_a&J\end{array}
\right\} \nonumber \\
&&\times G_{aa'bb'}(L)(I'n'||A^{\dag}_L(bb')||In).  \label{kkdo1.11} 
\end{eqnarray}
The normalization condition (\ref{kkp1.sum}) becomes
\begin{equation}
\sum_{aIn}[|v_{J\nu}(aIn)|^2 + |u_{J\nu}(aIn)|^2]=\Omega. \label{kkdo1.12}
\end{equation}

The equations derived above define a linear eigenvalue problem, provided we  supply from the outside the single-particle energies $h_a$, the reduced matrix elements of the included multipole and pairing forces, and the excitation energies
of the neighboring even nuclei.  In the underlying (self-consistent) theory these quantities, other than the single-particle energies, can themselves be expressed in terms of the CFP $v$ and $u$. In practice, characteristics of
even nuclei expressed   in terms of the reduced matrix elements of single-particle operators
\begin{equation}
F_{LM_L}=\sum_{ac}f_{ac}(L)B_{LM_L}(ac)  \label{sf1}
\end{equation}
and pair transfer operators
\begin{equation}
G_{LM_L}=\sum_{ab}\gamma _{ab}(L)A^{\dag}_{LM_L}(ab)  \label{sf2}
\end{equation}
are available rather than the reduced matrix elements of two-body interactions. To make use of them in the
equations (\ref{kkdo1.6}) and (\ref{kkdo1.7}) it is convenient to present the interactions appearing in Eqs. (\ref{kkdo1.8}) -- (\ref{kkdo1.11}) as a sum of separable interactions of the form:
\begin{eqnarray}
F_{acdb}(L)&=&-\kappa_L(q_aq_b)f_{ac}(L)f_{db}(L), \label{sf3}\\
G_{abcd}(L)&=&-g_L(q_a)\gamma_{ab}(L)\gamma_{cd}(L).  \label{sf4}
\end{eqnarray}
Then the interactions are parameterized by a few strengths $\kappa_L$ and $g_L$ which can be either fitted to the
experimental data or estimated theoretically.

\subsection{Physical solutions}

The equations that we have derived have the form of generalized Hartree-Bogoliubov (HB) equations.  We summarize the content of Eqs.~(\ref{kkdo1.6}) and (\ref{kkdo1.7}) in the condensed form
\begin{eqnarray}
{\cal H}\Psi_{J\mu\nu}&=&{\cal E}_{J\nu}\Psi_{J\mu\nu}, \label{kkdo1.13}\\
\Psi&=&\left(\begin{array}{c}v\\u \end{array}\right),   \label{kkdo1.14}\\
{\cal H}&=& \left(\begin{array}{cc}\epsilon'+\omega^{(A+1)}+\Gamma^{(A+1)}& 
\Delta \\ \Delta^{\dag} & -\epsilon''+\omega^{(A-1)}-\Gamma^{\dag(A-1)}\end{array}
\right). \label{kkdo1.15}
\end{eqnarray}

The HB structure of these equations implies that only half of the solutions refer to physical states.  In the standard ground-state problem, the solutions divide  into two sets with reversed energies, the positive energies representing the physical solutions.
The solutions of Eq.~(\ref{kkdo1.13}) do not divide so neatly.
The resolution of this dilemma starts by identifying a piece of the Hamiltonian ${\cal H}$ that has such a simple property and then initially to "turn off" the remainder of the operator. This is done with the aid of the orthogonal matrix $C$ that interchanges
particles and holes,
\begin{equation}
C=\left(\begin{array}{cc} 0&-1\\1&0 \end{array}\right) \label{kkdo1.16}
\end{equation} 
and its transpose $\tilde{C}$, and by defining the operator
\begin{eqnarray}
{\cal H}_o&=&\frac{1}{2}({\cal H}-C{\cal H}\tilde{C}) \nonumber \\
&& =\left(\begin{array}{cc}({\cal H}_o)_{11}&
\frac{1}{2}(\Delta+\Delta^{\dag}) \\
\frac{1}{2}(\Delta+\Delta^{\dag}) &-({\cal H}_o)_{11}
\end{array}\right), \label{kkdo1.17}\\
({\cal H}_o)_{11}&=&\frac{1}{2}(\epsilon'+\epsilon'')+\frac{1}{2}(\Gamma^{(A+1)}+\Gamma^{\dag(A-1)})
+\frac{1}{2}(\omega^{(A+1)}-\omega^{(A-1)}).    \label{kkdo1.18}
\end{eqnarray}
Because
\begin{equation}
C{\cal H}_o\tilde{C}=-{\cal H}_o,      \label{kkdo1.19}
\end{equation}
if $\Psi$ is an eigenstate of ${\cal H}_o$ with eigenvalue ${\cal E}$, then
$C\Psi$ is an eigenstate with eigenvalue $-{\cal E}$.  As in the simple case,
the solutions with positive eigenvalues are the physical solutions for our limiting case.

Next we turn on the remainder of the Hamiltonian, namely the even part
\begin{equation}
{\cal H}_e=\frac{1}{2}({\cal H} +C{\cal H}\tilde{C}),  \label{kkdo1.20}
\end{equation}
our aim being to keep track of the physical solutions.  In the applications carried out to date, we have described several methods for carrying out this program.  Initially we described methods based on turning on the "perturbation" slowly and following the
physical solutions by continuity arguments, but in the end realized \cite{pav4} that by invoking the no-crossing theorem for two levels of the same angular momentum, we can simply identify that half of the solutions of a given angular momentum with the
 largest energies as the physical
solutions.  From a practical point of view, it suffices to diagonalize the Hamiltonian
${\cal H}_e$ using the complete set of states ({\it physical and spurious}) generated by ${\cal H}_o$ and selecting the largest half of the eigenvalues as the physical solutions.
The diagonalizationof ${\cal H}_e$ within the subspace of physical (positive energy)
states of ${\cal H}_o$ performed originally when solving the model \cite{DF3} can lead to 
a bad approximation of physical solutions of ${\cal H}$ or even give some unphysical solutions, since matrix elements of ${\cal H}_e$ between physical and unphysical solutions
need not be small.

It is of interest to contrast this procedure with the one used earlier in which the term
${\cal H}_e$ was turned on adiabatically and the physical solutions followed by using a wave
function overlap argument.  This procedure is based on the assumption that the physical
wave functions change slowly during such a procedure.  It is precisely this assumption
that fails in the neighborhood of an avoided crossing, because when this occurs, it is
well-known that there is an interchange of wave functions between the two levels involved.
In other words, as opposed to the simple argument based on the ordering of the levels,
the set of wave functions assigned as physical must be modified as one passes a near
crossing.

This brings us to another issue that is both technical and physical.  The simplest
application of the KKDF method is to cases where there is well established band
structure, either rotational or vibrational, of the same type for both neighboring
even nuclei.  The problem is then to classify the states of the odd nucleus into bands.  For this case, the study initially of ${\cal H}_o$ can be useful.  This is because for the states belonging to the same band, states of different $J$ are practically 
degenerate, because of the smallness of $\omega^{(A+1)}-\omega^{(A-1)}$.  This was the method used in our early work \cite{pav1,pav2,pav3,pav4}.  For more complicated situations, we can identify different band members
by the structure of the states, in the sense that the expansion coefficients in terms of a given basis of states vary slowly with angular momentum \cite{pav7}.  Consistent with the identification by state vector, we should equally be able to associate
states into bands by calculating transition rates of a suitable collective operator, usually the electric quadrupole operator.

\subsection{Matrix elements of single-particle transition operators\label{tro}}

We complete the exposition of the general formalism for present purposes
by deriving formulas for transition amplitudes of a general (charge-conserving) one-body
operator.  We choose this operator to be a tensor of rank $L$,
$T_{LM_L}$, that we write in the form
\begin{equation}
T_{LM_L} =\sum_{\beta\gamma} t_{\beta\gamma}a_\beta^{\dag}a_\gamma.
\label{ppk1.Tensor}
\end{equation}
The notation is such that the quantities $t_{\alpha\beta}$ include a product
of matrix elements of single-particle operators and of associated coupling
strengths (charges, gyromagnetic ratios, etc.)
We wish to calculate the matrix element $\langle J'\mu'\nu'|T_{LM_L}|
J\mu\nu\rangle$.   To carry through the calculation, we substitute for the 
ket a formally exact expression in terms of the action of single-particle
operators on the states of the core \cite{DO,DF4,DF5}
\begin{eqnarray}
|J\mu\nu\rangle &=& \frac{1}{\Omega}\sum_{\alpha IMn}[U_{J\mu\nu}(\alpha
IMn)a_{\alpha}^{\dag}|\underline{IMn}\rangle   \nonumber \\
&& +V_{J\mu\nu}(\alpha IMn)a_{\bar{\alpha}}|\overline{IMn}\rangle],
\label{ppk1.statev}
\end{eqnarray}
where an underline identifies the lighter of the two cores and an overline 
the heavier one.  That Eq.~(\ref{ppk1.statev}) represents an orthonormal set can be proved by first showing that the orthogonality of different states follows from the equations of motion (\ref{ppk1.hfb1}) and (\ref{ppk1.hfb2}) and then showing that the 
normalization follows from the CFP normalization condition (\ref{kkp1.sum}).
By using the commutation relations and completeness, this leads to the 
following expression for the transition element:
\begin{eqnarray}                           
\langle J'\mu'\nu'|T_{LM_L}|J\mu\nu\rangle 
&=&\frac{1}{\Omega}\sum_{\alpha IMnI'M'n'}
[U_{J'\mu'\nu'}(\alpha I'M'n')U_{J\mu\nu}(\alpha IMn)  \nonumber \\
&&\times\langle \underline{I'M'n'}|T_{LM_L}|\underline{IMn}\rangle    
\nonumber \\
&&+ V_{J'\mu'\nu'}(\alpha I'M'n')V_{J\mu\nu}(\alpha IMn)
\langle \overline{I'M'n'}|T_{LM_L}|\overline{IMn}\rangle]    \nonumber \\
&&+\frac{1}{\Omega}\sum_{\alpha,\alpha',IMn}t_{\alpha\alpha'}[U_{J'\mu'\nu'}
(\alpha IMn)U_{J\mu\nu}(\alpha' IMn)  \nonumber \\
&& - V_{J\mu\nu}(\bar{\alpha} IMn)V_{J'\mu'\nu'}(\bar{\alpha}' IMn)].  \label{ppk1.tran}
\end{eqnarray}

This is now evaluated by use of the Wigner-Eckart theorem with the help of  Eqs.~ 
(\ref{kkdo1.1}) and (\ref{kkdo1.2}) and the following additional
definitions of  reduced matrix elements:
\begin{eqnarray}
&&\langle J'\mu'\nu'|T_{LM_L}|J\mu\nu\rangle =\frac{(-1)^{J-\mu}}
{\sqrt{2L+1}}(J'\mu'J-\mu|LM_L)  \nonumber\\
&& \times\langle J'\nu'||T_L||J\nu\rangle,\label{ppk1.T}
 \\
&&\langle I'M'n'|T_{LM_L}|IMn\rangle = \frac{(-1)^{I-M}}{\sqrt{2L+1}}
(I'M'I-M|LM_L)    \nonumber \\
&& \times\langle I'n'||T_L||In\rangle,  \label{ppk1.T1} \\
&&t_{\alpha\gamma} = \frac{(-1)^{j_c -m_c}}{\sqrt{2L+1}}
(j_a m_a j_c -m_c|LM_L) t_{ac}.        \label{ppk1.T2}
% &&V_{J\mu\nu}(\alpha,IMK)=\frac{(-1)^{J-\mu}}{\sqrt{2j_a +1}}
%(IMJ-\mu|j_a m_a)  v_{J\nu}(aIK),  \label{ppk1.WE1} \\
%&&U_{J\mu\nu}(\alpha,IMK)=\frac{(-1)^{J-\mu +j_a +m_a}}{\sqrt{2j_a +1}}
%(IMJ-\mu|j_a m_a) u_{J\nu}(aIK).  \label{ppk1.WE2}
\end{eqnarray}
We thus find
\begin{eqnarray}
&&\langle J'\nu'||T_L||J\nu\rangle = \frac{1}{\Omega}\sum_{aInI'n'}
(-1)^{j_a +J'+I+L}\left\{\begin{array}{ccc}I & I' & L \\ J' & J & j_a
\end{array}\right\}   \nonumber \\
&& \times \sqrt{(2J+1)(2J'+1)}[u_{J\nu}(aIn)u_{J'\nu'}(aI'n')\langle \underline{I'n'}||T_L||
\underline{In}\rangle  \nonumber \\
&& + v_{J\nu}(aIn)v_{J'\nu'}(aI'n')\langle \overline{I'n'}||T_L||
\overline{In}\rangle]  \nonumber \\
&&+\frac{1}{\Omega}\sum_{aa'In}t_{aa'}\sqrt{(2J+1)(2J'+1)}
[(-1)^{j_{a'} +I+J'+L}
\left\{\begin{array}{ccc}j_a & j_{a'} &L\\ J & J' & I
\end{array}\right\}u_{J'\nu'}(aIn)u_{J\nu}(a'In)  \nonumber \\
&&+(-1)^{j_{a'} +I+J'+1}\left\{\begin{array}{ccc}j_a & j_{a'} & L\\ J' & J & I
\end{array}\right\}v_{J\nu}(aIn)v_{J'\nu'}(a'In)].    \label{ppk1.RME}
\end{eqnarray}
We thus have a clear separation into collective and single-particle contributions.

\section{Odd-odd nuclei}

\subsection{Equations of motion}

We turn to the problem of deriving a general core-particle coupling model for
odd-odd nuclei analogous to the model derived for odd nuclei in Sec.~II.  Given an odd-odd nucleus with $Z$ protons and $N$ neutrons, we shall relate its properties to those of four neighboring even nuclei with proton-neutron numbers $(Z+1,N+1)$,
$(Z+1,N-1)$, $(Z-1,N+1)$ and $(Z-1,N-1)$ respectively.  In the following development,
we shall continue to use Greek letters for a general single-particle level, but
shall use $p$, $p'$, etc. to indicate proton levels and $n$, $n'$, etc. to indicate neutron levels.  To relate the properties of the target odd nucleus to its four even neighbors, we need the equations of motion for four pairs of operators that we present for initial convenience in an uncoupled form,
\begin{eqnarray}
[a_{\bar{p}}a_{\bar{n}},H]&=& (h_{p}'+h_{n}')a_{\bar{p}}a_{\bar{n}}\nonumber\\
&& +F_{\bar{p}\bar{p}'\beta\beta'}a_{\bar{p}'}a_{\bar{n}}(a_{\beta'}^{\dag}
a_{\beta}) +F_{\bar{n}\bar{n}'\beta\beta'}a_{\bar{p}}a_{\bar{n}'}(a_{\beta'}^{\dag}
a_{\beta}) \nonumber \\
&& +G_{\bar{p}p'p''p'''}a_{p'}^{\dag}a_{\bar{n}}(a_{p'''}a_{p''})
+G_{\bar{n}n'n''n'''}a_{\bar{p}}a_{n'}^{\dag}(a_{n'''}a_{n''}) \nonumber \\
&&-F_{\bar{p}\bar{p}'\bar{n}'\bar{n}}a_{\bar{p}'}a_{\bar{n}'},  \label{eomoo.1}\\
{}[a_{\bar{p}}a_{n}^{\dag},H]&=& (h_{p}'-h_{n}'')a_{\bar{p}}a_{n}^{\dag} \nonumber\\
&&+F_{\bar{p}\bar{p}'\beta\beta'}a_{\bar{p}'}a_{n}^{\dag}(a_{\beta'}^{\dag}
a_{\beta}) 
-F_{nn'\beta\beta'}a_{\bar{p}}a_{n'}^{\dag}(a_{\beta}^{\dag}
a_{\beta'}) \nonumber \\
&& +G_{\bar{p}p'p''p'''}a_{p'}^{\dag}a_{n}^{\dag}(a_{p'''}a_{p''})
-G_{n\bar{n}'n''n'''}a_{\bar{p}}a_{\bar{n}'}(a_{n''}^{\dag}a_{n'''}^{\dag}) \nonumber \\
&&+F_{\bar{p}\bar{p}'nn'}a_{\bar{p}'}a_{n'}^{\dag},  \label{eomoo.2}\\
{}[a_p^{\dag}a_{\bar{n}},H]&=& (-h_{p}''+h_{n}')a_p^{\dag}a_{\bar{n}}\nonumber\\
&&-F_{pp'\beta\beta'}a_{p'}^{\dag}a_{n}(a_{\beta}^{\dag}a_{\beta'}) 
+F_{\bar{n}\bar{n}'\beta\beta'}a_p^{\dag}a_{\bar{n}'}(a_{\beta'}^{\dag}
a_{\beta}) \nonumber \\
&&-G_{p\bar{p}'p''p'''}a_{\bar{p}'}a_{\bar{n}}(a_{p''}^{\dag}a_{p'''}^{\dag}) 
+G_{\bar{n}n'n''n'''}a^{\dag}_pa_{n'}^{\dag}(a_{n'''}a_{n''}) \nonumber \\
&&+F_{pp'\bar{n}\bar{n}'}a_{p'}^{\dag}a_{\bar{n}'},  \label{eomoo.3}\\
{}[a_p^{\dag}a_n^{\dag},H]&=& (-h_{p}''-h_{n}'')a_p^{\dag}a_n^{\dag}\nonumber\\
&&F_{pp'\beta\beta'}a_{p'}^{\dag}a_{n}^{\dag}(a_{\beta}^{\dag}a_{\beta'}) 
-F_{nn'\beta\beta'}a_p^{\dag}a_{n'}^{\dag}(a_{\beta}^{\dag}
a_{\beta'}) \nonumber \\
&&-G_{p\bar{p}'p''p'''}a_{\bar{p}'}a_n^{\dag}(a_{p''}^{\dag}a_{p'''}^{\dag}) 
-G_{n\bar{n}'n''n'''}a_p^{\dag}a_{\bar{n}'}(a_{n''}^{\dag}a_{n'''}^{\dag}) \nonumber \\
&&-F_{pp'n'n}a_{p'}^{\dag}a_{n'}^{\dag},  \label{eomoo.4}
\end{eqnarray}
Notice that we have not included neutron-proton pairing interactions.

We shall study matrix elements of these equations between the states $\langle JM_J s|$ on the left, an included state of the odd-odd nucleus (Z,N) and the appropriate one of the 
states $|\sigma\tau RM_R r\rangle$ of the even nucleus $(Z\sigma 1,N\tau 1)$, $\sigma=\pm$,
$\tau=\pm$.  For the further development of the formalism, in particular the reduction to equations for reduced matrix  elements, there are, however, several choices.  In this section, we shall develop a method that makes maximal use of the formalism for odd nuclei, and, as a consequence involves only single-particle CFP.
This method, which treats the neutron and proton asymmetrically, we shall refer to as sequential coupling.  In the next section, we shall develop an alternative, referred to as
symmetrical coupling, that bypasses any use of the results for odd nuclei.

\subsection{Equations of motion in sequential coupling}

By introducing a complete set of states of the appropriate odd nucleus between the neutron and proton single-particle operators, we write
\begin{eqnarray}
\langle JM_J s|a_{\bar{p}}a_{\bar{n}}|++IM_I r\rangle&=&\Psi^{(++)}_{JM_J s}(pnIM_I r)\nonumber \\
&=&\sum_{J_n M_n r_n}X_{JM_J s}(pJ_n M_n r_n)V^{(+)}_{J_N M_n r_n}(nIM_I r),  \label{seq.1} \\
\langle JM_J s|a_{\bar{p}}a^{\dag}_{n}|+-IM_I r\rangle&=&\Psi^{(+-)}_{JM_J s}(pnIM_I r)\nonumber\\
&=&\sum_{J_n M_n r_n}X_{JM_J s}(pJ_n M_n r_n)U^{(+)}_{J_N M_n r_n}(nIM_I r),  \label{seq.2} \\
\langle JM_J s|a^{\dag}_{p}a_{\bar{n}}|-+IM_I r\rangle&=&\Psi^{(-+)}_{JM_J s}(pnIM_I r)\nonumber \\
&=&\sum_{J_n M_n r_n}Y_{JM_J s}(pJ_n M_n r_n)V^{(-)}_{J_N M_n r_n}(nIM_I r),  \label{seq.3} \\
\langle JM_J s|a^{\dag}_{p}a^{\dag}_{n}|--IM_I r\rangle&=&\Psi^{(--)}_{JM_J s}(pnIM_I r)\nonumber \\
&=&\sum_{J_n M_n r_n}Y_{JM_J s}(pJ_n M_n r_n)U^{(-)}_{J_N M_n r_n}(NIM_I r),  \label{seq.4} 
\end{eqnarray}
Here
\begin{eqnarray}
V^{(\sigma)}_{J_n M_n r_n}(nIM_I r)&=&\langle\sigma J_n M_n r_n|a_{\bar{n}}|\sigma +IM_I r\rangle,   \label{seq.5}\\
U^{(\sigma)}_{J_n M_n r_n}(nIM_I r)&=&\langle\sigma J_n M_n r_n|a_{\bar{n}}|\sigma -IM_I r\rangle,   \label{seq.6}
\end{eqnarray}
are two sets of CFP amplitudes for odd neutron nuclei, which can be calculated using the formalism for odd nuclei developed in Sec.~II. On the other hand the amplitudes
\begin{eqnarray}
X_{JM_J s}(pJ_n M_n r_n)&=&\langle JM_J s|a_{\bar{p}}|+J_n M_n r_n\rangle,\label{seq.7}\\
Y_{JM_J s}(pJ_n M_n r_n)&=&\langle JM_J s|a^{\dag}_{p}|-J_n M_n r_n\rangle,\label{seq.8}
\end{eqnarray}
are single-particle CFP relating odd and odd-odd nuclei. The aim of the present coupling scheme is to obtain equations to determine the amplitudes $X$ and $Y$.  Before proceeding along these lines, we remark that there is a related sequential scheme obtained by starting with two-particle amplitudes in which the order of the single-particle operators is interchanged.

The next step is to write out equations for the amplitudes 
$\Psi^{(\sigma\tau)}_{JM_J s}(pnIM_I r)$.  To fix additional notation, which will be understood immediately to be a modified form of the notation of Sec.~II, we exhibit just one of these equations (using the summation convention),
\begin{eqnarray}
&&(-E_{Js}+E_{++Ir})\Psi^{(++)}_{JM_J s}(pnIM_I r)=(h'_p +h'_n)\Psi^{(++)}_{JM_J s}(pnIM_I r) \nonumber \\
&&+\Gamma^{(++)}(\bar{p}IM_I r|\bar{p}'I'M_{I'}r')\Psi^{(++)}_{JM_J s}(p'nI'M_{I'}r')
+\Delta^{(+)}(\bar{p}IM_I r|p'I'M_{I'}r')\Psi^{(-+)}_{JM_J s}(p'nI'M_{I'}r')\nonumber \\
&&+\Gamma^{(++)}(\bar{n}IM_I r|\bar{n}'I'M_{I'}r')\Psi^{(++)}_{JM_J s}(pn'I'M_{I'}r')
+\Delta^{(+)}(\bar{n}IM_I r|n'I'M_{I'}r')\Psi^{(+-)}_{JM_J s}(pn'I'M_{I'}r') \nonumber \\
&&-F_{\bar{p}\bar{p}'\bar{n}'\bar{n}}\Psi^{(++)}_{JM_J s}(p'n'IM_I r).  \label{seq.9}
\end{eqnarray}
Into this equation and its three partners, we substitute Eqs.~(\ref{seq.1})-(\ref{seq.4})
and recognize that the result can be simplified by the use of equations such as
\begin{eqnarray}
&&(E_{J_n r_n}-E_{++Ir}-h'_n)V^{(+)}_{J_n M_n r_N}(nIM_I r)   \nonumber\\
&&=\Gamma^{(++)}(\bar{n}IM_I r|\bar{n}'I'M_{I'}r')V^{(+)}_{J_n M_n r_n}(n'I'M_{I'}r')\nonumber\\
&&+\Delta^{(+)}(\bar{n}IM_I r|n'I'M_{I'}r')U^{(+)}_{J_n M-n r_n}(n'I'M_{I'}r') \label{seq.10}
\end{eqnarray}
and its partners.  The resulting equation for $\Psi^{(++)}$ is then combined with the corresponding equation for $\Psi^{(+-)}$ , contracting the first with a $V^{(+)}$ factor and the second with a $U^{(+)}$ factor to as to permit use of the normalization condition
(\ref{kkp1.sum}). We carry through a corresponding procedure for the pair of amplitudes
$\Psi^{(-+)}$ and $\Psi^{(--)}$.  

We thus obtain the pair of equations
\begin{eqnarray}
(-E_{Js}+E_{+J_n r_n})X_{JM_J s}(pJ_n M_n r_n)&=&h'_p X_{JM_J s}(pJ_n M_n r_n)\nonumber\\
&&+\Gamma^{(+)}(\bar{p}J_n M_n r_n|\bar{p}'J_{n'}M_{n'}r_{n'})X_{JM_J s}(p'J_{n'}M_{n'}r_{n'}) \nonumber \\ 
&&+\Delta(\bar{p}J_n M_n r_n|p'J_{n'}M_{n'}r_{n'})Y_{JM_J s}(p'J_{n'}M_{n'}r_{n'})\nonumber\\
&&+V^{(+)}(\bar{p}J_n M_n r_n|\bar{p}'J_{n'}M_{n'}r_{n'})X_{JM_J s}(p'J_{n'}M_{n'}r_{n'}), \label{seq.11}\\
(-E_{Js}+E_{-J_n r_n})Y_{JM_J s}(pJ_n M_n r_n)&=&-h''_p Y_{JM_J s}(pJ_n M_n r_n)\nonumber\\
&&-\Gamma^{\dag(-)}(pJ_n M_n r_n|p'J_{n'}M_{n'}r_{n'})Y_{JM_J s}(p'J_{n'}M_{n'}r_{n'})
\nonumber \\ 
&&-\Delta^{\dag}(pJ_n M_n r_n|\bar{p}'J_{n'}M_{n'}r_{n'})X_{JM_J s}(p'J_{n'}M_{n'}r_{n'})
\nonumber\\
&&+V^{(-)}(pJ_n M_n r_n|p'J_{n'}M_{n'}r_{n'})Y_{JM_J s}(p'J_{n'}M_{n'}r_{n'}). \label{seq.12}
\end{eqnarray}
Here
\begin{eqnarray}
&&\Gamma^{(+)}(pJ_n M_n r_n|p'J_{n'}M_{n'}r_{n'})=\sum_{nIM_I rI'M_{I'}r'}
\frac{1}{\Omega_{(n)}} \nonumber\\
&&\times [V^{(+)}_{J_n M_n r_n}(nIM_I r)\Gamma^{(++)}(pIM_I r|p'I'M_{I'}r')V^{(+)}_{J_{n'}M_{n'}r_{n'}}(nI'M_{I'}r') \nonumber\\
&&+U^{(+)}_{J_n M_n r_n}(nIM_I r)\Gamma^{(+-)}(pIM_I r|p'I'M_{I'}r')U^{(+)}_{J_{n'}M_{n'}r_{n'}}(nI'M_{I'}r')], \label{seq.13}\\
&&\Delta(pJ_n M_n r_n|p'J_{n'}M_{n'}r_{n'})=\sum_{nIM_I rI'M_{I'}r'}
\frac{1}{\Omega_{(n)}}\nonumber\\
&&\times [V^{(+)}_{J_n M_n r_n}(nIM_I r)\Delta^{(+)}(pIM_I r|p'I'M_{I'}r')V^{(-)}_{J_{n'}M_{n'}r_{n'}}(nI'M_{I'}r') \nonumber\\
&&+U^{(+)}_{J_n M_n r_n}(nIM_I r)\Delta^{(-)}(pIM_I r|p'I'M_{I'}r'}U^{(-)}_{J_{n'}M_{n'}r_{n'}(nI'M_{I'}r')], \label{seq.14} \\
&&V^{(+)}(\bar{p}I_n M_n r_n|\bar{p}'J_{n'}M_{n'}r_{n'})=\sum_{nIM_I rI'M_{I'}r'}
\frac{1}{\Omega_{(n)}}\nonumber\\
&&\times[-V^{(+)}_{J_n M_n r_n}(nIM_I r)V^{(+)}_{J_{n'}M_{n'}r_{n'}}(n'IM_{I}r) 
F_{\bar{p}\bar{p}'\bar{n}'\bar{n}} \nonumber\\
&&+U^{(+)}_{J_n M_n r_n}(nIM_I r) U^{(+)}_{J_{n'}M_{n'}r_{n'}}(n'IM_{I}r)F_{\bar{p}\bar{p}'nn'}],
\label{seq.15}\\
&&\Gamma^{\dag(-)}(pJ_n M_n r_n|p'J_{n'}M_{n'}r_{n'})=\sum_{nIM_I rI'M_{I'}r'}
\frac{1}{\Omega_{(n)}} \nonumber \\
&&\times [V^{(-)}_{J_n M_n r_n}(nIM_I r)\Gamma^{\dag(-+)}(pIM_I r|p'I'M_{I'}r')V^{(-)}_{J_{n'}M_{n'}r_{n'}}(nI'M_{I'}r') \nonumber\\
&&+U^{(-)}_{J_n M_n r_n}(nIM_I r)\Gamma^{\dag(--)}(pIM_I r|p'I'M_{I'}r'}U^{(-)}_{J_{n'}M_{n'}r_{n'}(nI'M_{I'}r')], \label{seq.16}\\
&&\Delta^{\dag}(pJ_n M_n r_n|p'J_{n'}M_{n'}r_{n'})=\sum_{nIM_I rI'M_{I'}r'}
\frac{1}{\Omega_{(n)}}\nonumber\\
&&\times [V^{(-)}_{J_n M_n r_n}(nIM_I r)\Delta^{\dag(+)}(pIM_I r|p'I'M_{I'}r'}V^{(+)}_{J_{n'}M_{n'}r_{n'}(nI'M_{I'}r') \nonumber\\
&&+U^{(-)}_{J_n M_n r_n}(nIM_I r)\Delta^{\dag(-)}(pIM_I r|p'I'M_{I'}r'}U^{(+)}_{J_{n'}M_{n'}r_{n'}(nI'M_{I'}r')], \label{seq.17} \\
&&V^{(-)}(pI_n M_n r_n|p'J_{n'}M_{n'}r_{n'})=\sum_{nIM_I rI'M_{I'}r'}
\frac{1}{\Omega_{(n)}}\nonumber\\
&&\times [V^{(-)}_{J_n M_n r_n}(nIM_Ir)V^{(-)}_{J_{n'}M_{n'}r_{n'}}(n'IM_{I}r) 
F_{pp'\bar{n}\bar{n}'} \nonumber\\
&&-U^{(-)}_{J_n M_n r_n}(nIM_I r)U^{(-)}_{J_{n'}M_{n'}r_{n'}}(n'IM_{I}r)F_{pp'nn'}].
\label{seq.18}
\end{eqnarray}

The correctness of the above equations can be verified independently.  By starting with
Eqs.~(\ref{ppk1.eom1}) and (\ref{ppk1.eom2}), we can derive equations of the form 
(\ref{ppk1.hfb1}) and (\ref{ppk1.hfb2}) with potentials $\Gamma$ and $\Delta$ that refer appropriately to the odd systems, rather than the even system and with no overt sign of the neutron-proton interaction terms.  These equations are readily transformed into the results
given above by the application of Eq.~(\ref{ppk1.statev}), just as the latter was applied in Sec.~II.D to express transition matrix elements between odd states in terms of matrix elements between even states and single-particle CFP.

The final goal of this section is to obtain equations of motion for the reduced matrix elements.  For this purpose the only definitions needed to supplement 
Eqs.~(\ref{kkdo1.1})-(\ref{kkdo1.4}) are those for the reduced CFP relating the odd to the odd-odd nuclei,
\begin{eqnarray}
X_{JM_J s}(PJ_n M_n r_n)=(-1)^{j_p -m_p}(J_n M_n j_p m_p|JM_J)\chi_{Js}(j_p J_n r_n),
\label{seq.19} \\
Y_{JM_J s}(PJ_n M_n r_n)=(J_n M_n j_p m_p|JM_J)\eta_{Js}(j_p J_n r_n). \label{seq.19a}
\end{eqnarray}
It is then a straightforward exercise in angular momentum algebra to derive the equations
\begin{eqnarray}
(-E_{Js}+E_{+J_n r_n})\chi_{Js}(j_p J_n r_n)&=& h'_p\chi_{Js}(j_p J_n r_n)\nonumber\\
&&+\Gamma^{(+)}(j_p J_n r_n|j_{p'}J_{n'}r_{n'})\chi_{Js}(j_{p'}J_{n'}r_{n'})\nonumber \\
&&+\Delta(j_p J_n r_n|j_{p'}J_{n'}r_{n'})\eta_{Js}(j_{p'}J_{n'}r_{n'}) \nonumber \\
&&+V^{(+)}(j_p J_n r_n|j_{p'}J_{n'}r_{n'})\chi_{Js}(j_{p'}J_{n'}r_{n'}),  \label{seq.20}\\
(-E_{Js}+E_{-J_n r_n})\eta_{Js}(j_p J_n r_n)&=& -h''_p\eta_{Js}(j_p J_n r_n)\nonumber\\
&&-\Gamma^{\dag(-)}(j_p J_n r_n|j_{p'}J_{n'}r_{n'})\eta_{Js}(j_{p'}J_{n'}r_{n'})\nonumber \\
&&+\Delta^{\dag}(j_p J_n r_n|j_{p'}J_{n'}r_{n'})\chi_{Js}(j_{p'}J_{n'}r_{n'}) \nonumber \\
&&+V^{(-)}(j_p J_n r_n|j_{p'}J_{n'}r_{n'})\eta_{Js}(j_{p'}J_{n'}r_{n'}),  \label{seq.21}
\end{eqnarray}
where
\begin{eqnarray}
\Gamma^{(+)}(j_p J_n r_n|j_{p'}J_{n'}r_{n'})&=&\frac{1}{\Omega_{(n)}}\left\{
\begin{array}{ccc}J_n&j_p&J\\j_{p'}&J_{n'}&L\end{array}\right\}
\left\{\begin{array}{ccc}I&J_n&j_n\\J_{n'}&I'&L\end{array}\right\}\nonumber \\
&&\times (-1`)^{j_p+j_n+J+I'+1-L}\sqrt{(2L+1)(2J_n+1)(2J_{n'}+1)(2I'+1)}\nonumber \\
&&\times F_{pp'aa'(L)}[v^{(+)}_{J_n r_n}(j_n Ir)v^{(+)}_{J_{n'}r_{n'}}(j_n I'r')
(I'r'||B_L^{(++)}(aa')||Ir)\nonumber\\
&&+u^{(+)}_{J_n r_n}(j_n Ir)u^{(+)}_{J_{n'}r_{n'}}(j_n I'r')
(I'r'||B_L^{(+-)}(aa')||Ir)],\label{seq.22}\\
\Delta(j_p J_n r_n|j_{p'}J_{n'}r_{n'})&=&\frac{1}{\Omega_{(n)}}\left\{
\begin{array}{ccc}J_n&j_p&J\\j_{p'}&J_{n'}&L\end{array}\right\}
\left\{\begin{array}{ccc}I&J_n&j_n\\J_{n'}&I'&L\end{array}\right\}\nonumber \\
&&\times (-1)^{j_p+j_n+J+I'-L}\sqrt{(2L+1)(2J_n+1)(2J_{n'}+1)(2I'+1)}\nonumber \\
&&\times G_{pp'aa'(L)}[v^{(+)}_{J_n r_n}(j_n Ir)v^{(-)}_{J_{n'}r_{n'}}(j_n I'r')
(I'r'||A_L^{(+)}(aa')||Ir)\nonumber \\
&&+u^{(+)}_{J_n r_n}(j_n Ir)u^{(-)}_{J_{n'}r_{n'}}(j_n I'r')
(I'r'||A_L^{(-)}(aa')||Ir)],\label{seq.23}\\
V^{(+)}(j_p J_n r_n|j_{p'}J_{n'}r_{n'})&=& \frac{1}{\Omega_{(n)}}\left\{
\begin{array}{ccc}J_n&j_p&J\\j_{p'}&J_{n'}&L\end{array}\right\}
\left\{\begin{array}{ccc}j_{n'}&j_n&L\\J_n&J_{n'}&I\end{array}\right\}\nonumber \\
&&\times(-1)^{j_p+j_n+J_n+J+L}(2L+1)\sqrt{(2J_n+1)(2J_{n'}+1)}\nonumber\\
&&\times[v^{(+)}_{J_n r_n}(j_n Ir)v^{(+)}_{J_{n'}r_{n'}}(j_{n'}Ir)F_{pp'n'n}(L) \nonumber\\
&&+(-1)^{j_n+j_{n'}-L}u^{(+)}_{J_n r_n}(j_n Ir)u^{(+)}_{J_{n'}r_{n'}}(j_{n'}Ir)
 F_{pp'nn'}(L)], \label{seq.24}\\
\Gamma^{\dag(-)}(j_p J_n r_n|j_{p'}J_{n'}r_{n'})&=&\frac{1}{\Omega_{(n)}}\left\{
\begin{array}{ccc}J_n&j_p&J\\j_{p'}&J_{n'}&L\end{array}\right\}
\left\{\begin{array}{ccc}I&J_n&j_n\\J_{n'}&I'&L\end{array}\right\}\nonumber \\
&&\times (-1`)^{j_p+j_n+J+I'+1-L}\sqrt{(2L+1)(2J_n+1)(2J_{n'}+1)(2I'+1)}\nonumber \\
&&\times F_{pp'aa'}[v^{(-)}_{J_n r_n}(j_n Ir)v^{(-)}_{J_{n'}r_{n'}}(j_n I'r')
(I'r'||B_L^{(++)}(aa')||Ir)\nonumber\\
&&+u^{(+)}_{J_n r_n}(j_n Ir)u^{(+)}_{J_{n'}r_{n'}}(j_n I'r')
(I'r'||B_L^{(+-)}(aa')||Ir)],\label{seq.25}\\
\Delta^{\dag}(j_p J_n r_n|j_{p'}J_{n'}r_{n'})&=&\frac{1}{\Omega_{(n)}}\left\{
\begin{array}{ccc}J_n&j_p&J\\j_{p'}&J_{n'}&L\end{array}\right\}
\left\{\begin{array}{ccc}I&J_n&j_n\\J_{n'}&I'&L\end{array}\right\}\nonumber \\
&&\times (-1)^{j_p+j_n+J+I'-L}\sqrt{(2L+1)(2J_n+1)(2J_{n'}+1)(2I'+1)}\nonumber \\
&&\times G_{pp'aa'}[v^{(-)}_{J_n r_n}(j_n Ir)v^{(+)}_{J_{n'}r_{n'}}(j_n I'r')
(I'r'||A_L^{\dag(+)}(aa')||Ir)\nonumber \\
&&+u^{(-)}_{J_n r_n}(j_n Ir)u^{(+)}_{J_{n'}r_{n'}}(j_n I'r')
(I'r'||A_L^{\dag(-)}(aa')||Ir)],\label{seq.26}\\
V^{(-)}(j_p J_n r_n|j_{p'}J_{n'}r_{n'})&=& \frac{1}{\Omega_{(n)}}\left\{
\begin{array}{ccc}J_n&j_p&J\\j_{p'}&J_{n'}&L\end{array}\right\}
\left\{\begin{array}{ccc}j_{n'}&j_n&L\\J_n&J_{n'}&I\end{array}\right\}\nonumber \\
&&\times(-1)^{j_p+j_{n'}+J_n+J+1}(2L+1)\sqrt{(2J_n+1)(2J_{n'}+1)}\nonumber \\
&&\times[v^{(-)}_{J_n r_n}(j_n Ir)v^{(-)}_{J_{n'}r_{n'}}(j_{n'}Ir)F_{pp'nn'}(L) \nonumber\\
&&+(-1)^{j_n+j_{n'}-L}u^{(-)}_{J_n r_n}(j_n Ir)u^{(-)}_{J_{n'}r_{n'}}(j_{n'}Ir)
 F_{pp'n'n}(L)]. \label{seq.27}
 \end{eqnarray}

The normalization condition associated with this formalism is
\begin{equation}
\sum_{j_p J-n r_n}[|\chi_{Js}(j_p J_n r_n)|^2 +|\eta_{Js}(j_p J_n r_n)|^2]=\Omega_{(p)}.
\label{seq.27a}
\end{equation}

\subsection{Physical solutions}

The problem of choosing the physical solutions of Eqs.~(\ref{seq.20}) and (\ref{seq.21})
can solved by simply repeating the arguments given in Sec.~II.C.  This is seen immediately if we rearrange the energies in these equations so that they resemble exactly the corresponding equations (\ref{kkdo1.6}) and (\ref{kkdo1.7}). We thus write
\begin{eqnarray}
{\cal E}_{Js}\chi_{Js}(j_p J_n r_n)&=& (\epsilon'_p-\omega_{+J_n r_n})\chi_{Js}(j_p J_n r_n)\nonumber\\
&&+\Gamma^{(+)}(j_p J_n r_n|j_{p'}J_{n'}r_{n'})\chi_{Js}(j_{p'}J_{n'}r_{n'})\nonumber \\
&&+\Delta(j_p J_n r_n|j_{p'}J_{n'}r_{n'})\eta_{Js}(j_{p'}J_{n'}r_{n'}) \nonumber \\
&&+V^{(+)}(j_p J_n r_n|j_{p'}J_{n'}r_{n'})\chi_{Js}(j_{p'}J_{n'}r_{n'}),  \label{seq.28}\\
{\cal E}_{Js}\eta_{Js}(j_p J_n r_n)&=& (-\epsilon''_p-\omega_{-J_n r_n})\eta_{Js}(j_p J_n r_n)\nonumber\\
&&-\Gamma^{\dag(-)}(j_p J_n r_n|j_{p'}J_{n'}r_{n'})\eta_{Js}(j_{p'}J_{n'}r_{n'})\nonumber \\
&&+\Delta^{\dag}(j_p J_n r_n|j_{p'}J_{n'}r_{n'})\chi_{Js}(j_{p'}J_{n'}r_{n'}) \nonumber \\
&&+V^{(-)}(j_p J_n r_n|j_{p'}J_{n'}r_{n'})\eta_{Js}(j_{p'}J_{n'}r_{n'}),  \label{seq.29}
\end{eqnarray}
with
\begin{eqnarray}
{\cal E}_{Js}&=&-E_{Js}+\frac{1}{2}(E_+ +E_-),  \label{seq.30} \\
\epsilon&=& h-\frac{1}{2}(E_+ -E_-),  \label{seq.31} \\
\omega_{\pm J_n r_n}&=&E_{\pm J_n r_n}-E_{\pm}.   \label{seq.32}
\end{eqnarray}
Here $E_{\pm}$ are the ground states of the heavier and lighter odd neutron nuclei, respectively.
With these forms one has an exact parallel to Eqs.~(\ref{kkdo1.6}) and (\ref{kkdo1.7}),
and thus the arguments for choosing physical solutions can be repeated without modification.

\subsection{Matrix elements of single-particle transition operators}

The result we want can be read off directly from Eq.~(\ref{ppk1.RME}) if we replace, appropriately, the single-particle CFP $v$ and $u$ by $\chi$ and $\eta$ and the reduced
matrix elements of the transition operator $T_L$ between states of even nuclei by the corresponding matrix elements between states of the appropriate odd nuclei.  We thus obtain
\begin{eqnarray}
&&\langle J's'||T_L||Js\rangle=\frac{1}{\Omega_{(n)}}\sum_{j_p J_n r_n J_{n'}r_{n'}}
(-1)^{j_p+J'+J_n+L}\left\{\begin{array}{ccc}J_n&J_{n'}&L\\J'&J&j_p\end{array}\right\}
\nonumber\\
&&\times\sqrt{(2J+1)(2J'+1)}[\eta_{Js}(j_p J_n r_n)\eta_{J's'}(j_p J_{n'}r_{n'})
\langle\underline{J_{n'}r_{n'}}||T_L||\underline{J_n r_n}\rangle \nonumber \\
&&+\chi_{Js}(j_p J_n r_n)\chi_{J's'}(j_p J_{n'}r_{n'})
\langle\overline{J_{n'}r_{n'}}||T_L||\overline{J_n r_n}\rangle] \nonumber \\
&&+\frac{1}{\Omega_{n}}\sum_{j_p j_{p'}J_n r_n}t_{pp'}\sqrt{(2J+1)(2J'+1)}
[(-1)^{j_{p'}+J_n +J'+L}\left\{\begin{array}{ccc}j_p&j_{p'}&L\\J&J'&J_n\end{array}\right\}
\eta_{J's'}(j_p J_n r_n)\eta_{Js}(j_{p'}J_n r_n) \nonumber\\
&&+(-1)^{j_{p'}+J_n+J'+1}\left\{\begin{array}{ccc}j_n&j_{n'}&L\\J'&J&J_n\end{array}\right\}
\chi_{Js}(j_p J_n r_n)\chi_{J's'}(j_{p'}J_n r_n).   \label{seq.33}
\end{eqnarray}

\section{Formalism for symmetrical treatment of odd nucleons} 

\subsection{Equations for reduced matrix elements}

With the aid of the Wigner-Eckart theorem and suitable definitions, we proceed 
to the transformation of these raw equations to equations for reduced matrix elements. First we introduce a two-component operator $a_p^\sigma$ for protons,
\begin{eqnarray}
a_p^\sigma&=&a_{\bar{p}}(-1)^{j_p-m_p},\;\;\sigma=+,  \nonumber \\
          &=&a_p^{\dag},\;\;\sigma=-,    \label{eomoo.5}
\end{eqnarray}
and a corresponding pair, $a_n^\tau$, $\tau=\pm$, for neutrons.  We then couple the products,
\begin{equation}
a_p^\sigma a_n^\tau = \sum_l (j_p m_p j_n m_n|lm){\cal B}_{lm}^{\sigma\tau}(pn),
\label{eomoo.6}
\end{equation}
and define reduced matrix elements $b_{Js}^{\sigma\tau}(pnlRr)$,
\begin{equation}
\langle JM_J s|{\cal B}_{lm}^{\sigma\tau}(pn)|\sigma\tau RM_R r\rangle
= (RM_r lm|JM_j)b_{Js}^{\sigma\tau}(pnlRr). \label{eomoo.7} 
\end{equation}
Here $|JM_J s\rangle$ is a state of the odd-odd nucleus $(Z,N)$ and 
$|\sigma\tau RM_R r\rangle$ is a state of the even nucleus $(Z\sigma 1,N\tau 1)$.

In the following, we also require reduced matrix elements of the multipole and pairing operators, defined as follows,
\begin{eqnarray}
\langle\sigma\tau R'M_{R'}r'|B_{LM_L}(bb')|\sigma\tau RM_R r\rangle
&=&(-1)^{L-M_L}(RM_R L-M_L|R'M_{R'})\nonumber \\ &&\times (R'r'||B_L^{(\sigma\tau)}(bb')||Rr), \label{eomoo.8} \\
\langle\sigma\tau R'M_{R'}r'|B^{\dag}_{LM_L}(bb')|\sigma\tau RM_R r\rangle
&=& (RM_R LM_L|R'M_{R'})(R'r'||B_L^{\dag(\sigma\tau)}(bb')||Rr), \label{eomoo.8a} \\
\langle\sigma -R'M_{R'}r'|A_{LM}(nn')|\sigma + RM_R r\rangle 
&=&(-1)^{L-M_L}(RM_R L-M_L|R'M_{R'})\nonumber \\ &&\times(R'r'||A_L^{(\sigma)}(nn')||Rr),  \label{eomoo.9} \\
\langle\sigma +R'M_{R'}r'|A^{\dag}_{LM}(nn')|\sigma - RM_R r\rangle 
&=& (RM_R LM_L|R'M_{R'})(R'r'||A_L^{\dag(\sigma)}(nn')||Rr),  \label{eomoo.9a} \\
\langle-\tau R'M_{R'}r'|A_{LM}(pp')|+\tau RM_R r\rangle 
&=&(-1)^{L-M_L}(RM_R L-M_L|R'M_{R'})\nonumber\\ &&\times(R'r'||A_L^{(\tau)}(pp')||Rr),  \label{eomoo.10} \\
\langle+\tau R'M_{R'}r'|A^{\dag}_{LM}(pp')|-\tau RM_R r\rangle 
&=& (RM_R LM_L|R'M_{R'})(R'r'||A_L^{\dag(\tau)}(pp')||Rr).  \label{eomoo.10a} 
\end{eqnarray}
In the final equations of motion given below, we also introduce in as close analogy as possible with our procedure for the odd-nucleus case, various combinations of energies.  The energies of the odd states will naturally be specified by $E_{Js}$, those of the four neighboring even nuclei by 
$E_{\sigma\tau Rr}$, the ground states of the latter by $E_{\sigma\tau}$.  We then introduce the following differences
\begin{eqnarray}
{\cal E}_{Js}&=& -E_{Js}+\frac{1}{4}(E_{++}+E_{+-}+E_{-+}+E_{--}),
\label{eomoo.11}\\
\omega_{\sigma\tau Rr}&=& E_{\sigma\tau Rr}-E_{\sigma\tau}, \label{eomoo.12}\\
\epsilon_n^{\prime}&=& h_n^{\prime}-\frac{1}{4}(E_{++}-E_{+-})-\frac{1}{8}
(E_{++}-E_{--}), \label{eomoo.13} \\
\epsilon_p^{\prime}&=& h_p^{\prime}-\frac{1}{4}(E_{++}-E_{-+})-\frac{1}{8}
(E_{++}-E_{--}), \label{eomoo.14} \\
\epsilon_n^{\prime\prime}&=& h_n^{\prime\prime}-\frac{3}{4}(E_{++} 
-E_{+-})+\frac{1}{8}(E_{++}-E_{--}), \label{eomoo.15} \\
\epsilon_p^{\prime\prime}&=& h_p^{\prime\prime}-\frac{3}{4}(E_{++} 
-E_{-+})+\frac{1}{8}(E_{++}-E_{--}), \label{eomoo.16} \\
\epsilon_n^{\prime\prime\prime}&=& h_n^{\prime\prime}+\frac{1}{4}(E_{--} 
-E_{+-})+\frac{1}{8}(E_{--}-E_{++}), \label{eomoo.17} \\
\epsilon_p^{\prime\prime\prime}&=& h_p^{\prime\prime}+\frac{1}{4}(E_{--} 
-E_{-+})+\frac{1}{8}(E_{--}-E_{++}). \label{eomoo.18} 
\end{eqnarray}
With the included matrix elements defined below, we thereby obtain the following equations of motion for the reduced matrix elements,
\begin{eqnarray}
{\cal E}_{Js}b^{\sigma\tau}_{Js}(pnlRr)=\sum_{\sigma'\tau'p'n'l'R'r'}
{\cal H}(\sigma\tau pnlRr|\sigma'\tau' p'n'l'R'r')b^{\sigma'\tau'}_{Js}
(p'n'l'R'r'),   \label{eomoo.19}
\end{eqnarray}
with values for the non-vanishing matrix elements of the effective Hamiltonian ${\cal H}$ given by the expressions
\begin{eqnarray}
{\cal H}(++pnlRr|++p'n'l'R'r')&=&(\epsilon_p^{\prime}+\epsilon_n^{\prime}
-\omega_{++Rr})\delta_{p,p'}\delta_{n,n'}\delta_{l,l'}\delta_{R,R'}\delta_{r,r'}
\nonumber\\
&&+\Gamma_J^{(++)}(plRr|p'l'R'r')\delta_{n,n'}
+\Gamma_J^{(++)}(nlRr|n'l'R'r')\delta_{p,p'}  \nonumber\\
&&+V^{(++)}(pn|p'n')\delta_{l,l'}\delta_{R,R'}\delta_{r,r'},  \label{eomoo.20}\\
{\cal H}(+-pnlRr|+-p'n'l'R'r')&=&(\epsilon_p^{\prime}-\epsilon_n^{\prime\prime}
-\omega_{+-Rr})\delta_{p,p'}\delta_{n,n'}\delta_{l,l'}\delta_{R,R'}\delta_{r,r'}
\nonumber\\
&&+\Gamma_J^{(+-)}(plRr|p'l'R'r')\delta_{n,n'}
-\Gamma_J^{\dag(+-)}(nlRr|n'l'R'r')\delta_{p,p'}  \nonumber\\
&&+V^{(+-)}(pn|p'n')\delta_{l,l'}\delta_{R,R'}\delta_{r,r'},  \label{eomoo.21}\\
{\cal H}(-+pnlRr|-+p'n'l'R'r')&=&(-\epsilon_p^{\prime\prime}+\epsilon_n^{\prime}
-\omega_{-+Rr})\delta_{p,p'}\delta_{n,n'}\delta_{l,l'}\delta_{R,R'}\delta_{r,r'}
\nonumber\\
&&-\Gamma_J^{\dag(-+)}(plRr|p'l'R'r')\delta_{n,n'}
+\Gamma_J^{(-+)}(nlRr|n'l'R'r')\delta_{p,p'}  \nonumber\\
&&+V^{(-+)}(pn|p'n')\delta_{l,l'}\delta_{R,R'}\delta_{r,r'},  \label{eomoo.22}\\
{\cal H}(--pnlRr|--p'n'l'R'r')&=&(-\epsilon_p^{\prime\prime\prime}-\epsilon_n^{\prime\prime\prime}-\omega_{--Rr})\delta_{p,p'}\delta_{n,n'} \delta_{l,l'}\delta_{R,R'}\delta_{r,r'}
\nonumber\\
&&-\Gamma_J^{\dag(--)}(plRr|p'l'R'r')\delta_{n,n'}
-\Gamma_J^{\dag(--)}(nlRr|n'l'R'r')\delta_{p,p'}  \nonumber\\
&&+V^{(--)}(pn|p'n')\delta_{l,l'}\delta_{R,R'}\delta_{r,r'},  \label{eomoo.23}\\
{\cal H}(++pnlRr|-+p'n'l'R'r')&=& \Delta_J^{(+)}(plRr|p'l'R'r')\delta_{n,n'},
\label{eomoo.24}\\
{\cal H}(++pnlRr|+-p'n'l'R'r')&=& \Delta_J^{(+)}(nlRr|n'l'R'r')\delta_{p,p'},
\label{eomoo.25}\\
{\cal H}(+-pnlRr|--p'n'l'R'r')&=& \Delta_J^{(-)}(plRr|p'l'R'r')\delta_{n,n'},
\label{eomoo.26}\\
{\cal H}(+-pnlRr|++p'n'l'R'r')&=& \Delta_J^{\dag(+)}(nlRr|n'l'R'r')\delta_{p,p'},
\label{eomoo.27}\\
{\cal H}(-+pnlRr|++p'n'l'R'r')&=&\Delta_J^{\dag(-)}(plRr|p'l'R'r')\delta_{n,n'},
\label{eomoo.28}\\
{\cal H}(-+pnlRr|--p'n'l'R'r')&=& \Delta_J^{(-)}(nlRr|n'l'R'r')\delta_{p,p'},
\label{eomoo.29}\\
{\cal H}(--pnlRr|+-p'n'l'R'r')&=& \Delta_J^{\dag(-)}(plRr|p'l'R'r')\delta_{n,n'},
\label{eomoo.30}\\
{\cal H}(--pnlRr|-+p'n'l'R'r')&=& \Delta_J^{\dag(-)}(nlRr|n'l'R'r')\delta_{p,p'}.
\label{eomoo.31}
\end{eqnarray}
The remaining matrix elements ${\cal H}(\sigma\tau|-\sigma -\tau)$ vanish.

The effective interactions that occur in the above equations are
\begin{eqnarray}
\Gamma_J^{(\sigma\tau)}(plRr|p'l'R'r')&=&\sum_{Lbb'}(-1)^{j_{p}+j_n+l+l'+L+R+J}
\nonumber\\
&&\times\sqrt{(2R'+1)(2L+1)(2l+1)(2l'+1)}\left\{\begin{array}{ccc}
l&R&J\\R'&l'&L\end{array}\right\}\left\{\begin{array}{ccc}j_n&j_p&l\\L&l'&j_{p'}
\end{array}\right\}\nonumber\\
&&\times\bar{F}_{pp'bb'}(L)(R'r'||B_L^{(\sigma\tau)}(bb')||Rr),  \label{eomoo.32}\\
\Gamma_J^{(\sigma\tau)}(nlRr|n'l'R'r')&=&\sum_{Lbb'}(-1)^{j_{p}+j_{n'}+L+R+J}
\nonumber\\
&&\times\sqrt{(2R'+1)(2L+1)(2l+1)(2l'+1)}\left\{\begin{array}{ccc}
l&R&J\\R'&l'&L\end{array}\right\}\left\{\begin{array}{ccc}j_p&j_n&l\\L&l'&j_{n'}
\end{array}\right\}\nonumber\\
&&\times\bar{F}_{nn'bb'}(L)(R'r'||B_L^{(\sigma\tau)}(bb')||Rr),  \label{eomoo.33}\\
\Gamma_J^{\dag(\sigma\tau)}(plRr|p'l'R'r')&=&\sum_{Lbb'}(-1)^{j_{p}+j_{n}+L
+l+l'+R+J} \nonumber\\
&&\times\sqrt{(2R'+1)(2L+1)(2l+1)(2l'+1)}\left\{\begin{array}{ccc}
l&R&J\\R'&l'&L\end{array}\right\}\left\{\begin{array}{ccc}j_n&j_{p}&l\\L&l'&j_{p'}
\end{array}\right\}\nonumber\\
&&\times\bar{F}_{pp'bb'}(L)(R'r'||B^{\dag(\sigma\tau)}L^{\sigma}(bb')||Rr),  \label{eomoo.34}\\
\Gamma_J^{\dag(\sigma\tau)}(nlRr|n'l'R'r')&=&\sum_{Lbb'}(-1)^{j_{p}+j_{n'}+L+R+J} \nonumber\\
&&\times\sqrt{(2R'+1)(2L+1)(2l+1)(2l'+1)}\left\{\begin{array}{ccc}
l&R&J\\R'&l'&L\end{array}\right\}\left\{\begin{array}{ccc}j_p&j_{n}&l\\L&l'&j_{n'}
\end{array}\right\}\nonumber\\
&&\times\bar{F}_{nn'bb'}(L)(R'r'||B_L^{\dag(\sigma\tau)}(bb')||Rr),  \label{eomoo.35}\\
\Delta_J^{(\tau)}(plRr|p'l'R'r')&=&-\sum_{Lp''p'''}(-1)^{j_{p}+j_n+l+l'+L+R+J}
\nonumber\\
&&\times\sqrt{(2R'+1)(2L+1)(2l+1)(2l'+1)}\left\{\begin{array}{ccc}
l&R&J\\R'&l'&L\end{array}\right\}\left\{\begin{array}{ccc}j_n&j_p&l\\L&l'&j_{p'}
\end{array}\right\}\nonumber\\
&&\times G_{pp'p''p'''}(L)(R'r'||A_L^{(\tau)}(p''p''')||Rr),  \label{eomoo.36}\\
\Delta_J^{(\sigma)}(nlRr|n'l'R'r')&=&-\sum_{Ln''n'''}(-1)^{j_{p}+j_{n'}+L+R+J}
\nonumber\\
&&\times\sqrt{(2R'+1)(2L+1)(2l+1)(2l'+1)}\left\{\begin{array}{ccc}
l&R&J\\R'&l'&L\end{array}\right\}\left\{\begin{array}{ccc}j_p&j_n&l\\L&l'&j_{n'}
\end{array}\right\}\nonumber\\
&&\times G_{nn'n''n'''}(L)(R'r'||A_L^{(\sigma)}(n''n''')||Rr),  \label{eomoo.37}\\
\Delta_J^{\dag(\tau)}(plRr|p'l'R'r')&=&-\sum_{Lp''p'''}(-1)^{j_{p}+j_{n}+L+l+l'
+R+J}    \nonumber\\
&&\times\sqrt{(2R'+1)(2L+1)(2l+1)(2l'+1)}\left\{\begin{array}{ccc}
l&R&J\\R'&l'&L\end{array}\right\}\left\{\begin{array}{ccc}j_n&j_{p}&l\\L&l'&j_{p'}
\end{array}\right\}\nonumber\\
&&\times G_{pp'p''p'''}(L)(R'r'||A_L^{\dag(\tau)}(p''p''')||Rr),  \label{eomoo.38}\\
\Delta_J^{\dag(\sigma)}(nlRr|n'l'R'r')&=-&\sum_{Ln''n'''}(-1)^{j_{p}+j_{n'}+L+R+J}
\nonumber\\
&&\times\sqrt{(2R'+1)(2L+1)(2l+1)(2l'+1)  }\left\{\begin{array}{ccc}
l&R&J\\R'&l'&L\end{array}\right\}\left\{\begin{array}{ccc}j_p&j_{n}&l\\L&l'&j_{n'}
\end{array}\right\}\nonumber\\
&&\times G_{nn'n''n'''}(L)(R'r'||A_L^{\dag(\sigma)}(n''n''')||Rr),  \label{eomoo.39}\\
V^{(++)}(pn|p'n')&=-&\sum_L (-1)^{j_{p }+j_{n'}+l}(2L+1)
\nonumber\\
&&\times\left\{\begin{array}{ccc}j_p&j_n&l\\j_{n'}&j_{p'}&L\end{array}\right\}
\bar{F}_{pp'n'n}(L), \label{eomoo.40} \\
V^{(+-)}(pn|p'n')&=&\sum_L (-1)^{j_{p}+j_{n}+l+L}(2L+1)
\nonumber\\
&&\times\left\{\begin{array}{ccc}j_p&j_n&l\\j_{n'}&j_{p'}&L\end{array}\right\}
\bar{F}_{pp'nn'}(L), \label{eomoo.41} \\
V^{(-+)}(pn|p'n')&=&\sum_L (-1)^{j_{p'}+j_{n'}+l+L}(2L+1)
\nonumber\\
&&\times\left\{\begin{array}{ccc}j_p&j_n&l\\j_{n'}&j_{p'}&L\end{array}\right\}
\bar{F}_{pp'nn'}(L), \label{eomoo.42} \\
V^{(--)}(pn|p'n')&=&-\sum_L (-1)^{j_{p'}+j_{n}+l}(2L+1)
\nonumber\\
&&\times\left\{\begin{array}{ccc}j_p&j_n&l\\j_{n'}&j_{p'}&L\end{array}\right\}
\bar{F}_{pp'n'n}(L). \label{eomoo.42a} 
\end{eqnarray}

To the equations of motion, we add a normalization condition that can be derived from the anticommutation relation
\begin{eqnarray}
\sum_{pn}\{a_p,a_p^{\dag}\}\{a_n,a_n^{\dag}\}&=&\Omega_{(p)}\Omega_{(n)},
\label{eomoo.43}\\
\Omega_{(p)}&=&\sum_{p}(2j_p+1). \label{eomoo.44}
\end{eqnarray}
Rearranging the order of the operators, taking a diagonal matrix element in the state $|JM_J s\rangle$, utilizing completeness and the definitions (\ref{eomoo.6})
and (\ref{eomoo.7}), we obtain the expected result 
\begin{eqnarray}
\sum_{\sigma\tau j_p j_n lRr}|b_{Js}^{(\sigma\tau)}(pnlRr)|^2&=&
\Omega_{(p)}\Omega_{(n)}.  \label{eomoo.45}
\end{eqnarray}

\subsection{Physical solutions}

We expect the space of physical solutions to be only a quarter of the total space of solutions. With a little care, we can generalize the method used to identify physical solutions for the case of odd nuclei.  If we examine the Hamiltonian matrix 
${\cal H}$ given by Eqs.~(\ref{eomoo.20})-(\ref{eomoo.31}), 
we see that it can be decomposed into a sum
\begin{equation}
{\cal H}={\cal H}_p+{\cal H}_n -\omega +V_{np},  \label{eomoo.46}
\end{equation}
describing in turn an odd-neutron nucleus, an odd-proton nucleus, an excitation energy matrix, and a neutron-proton interaction energy.  We initially turn off the last two terms. Next we define two four-by-four matrices, $C_{n}$ and $C_{p}$,
\begin{eqnarray}
C_{n}&=&\left(\begin{array}{cc}\underline{C}&\underline{0}\\ \underline{0}&
\underline{C}\end{array}\right),  \label{eomoo.47} \\
C_{p}&=&\left(\begin{array}{cc}\underline{0}&-\underline{1}\\ \underline{1}&
\underline{0}\end{array}\right),  \label{eomoo.48}
\end{eqnarray}
where the underlined entries are each two-by-two matrices and the matrix $\underline{C}$ is 
the particle-hole conjugation matrix defined in Eq.~(\ref{kkdo1.16}).  The matrices $C_p$, $C_n$ commute with each other.

We then observe that the averages
\begin{eqnarray}
\bar{\cal H}_p&=&\frac{1}{2}({\cal H}_p +C_n{\cal H}_p\tilde{C}_n), \label{eomoo.49}\\
\bar{\cal H}_n&=&\frac{1}{2}({\cal H}_n +C_p{\cal H}_n\tilde{C}_p) 
\label{eomoo.50}
\end{eqnarray}
each have a structure more symmetrical than their individual terms.  Thus, the non-vanishing elements of $\bar{\cal H}_p$ are (in a condensed notation)
\begin{eqnarray}
\bar{\cal H}_p(++|++)&=&\bar{\cal H}_p(+-|+-)  \nonumber\\
&=&\epsilon_p^{\prime}+\frac{1}{2}(\Gamma_p^{(++)}+\Gamma_p^{(+-)}) ,   \label{eomoo.51} \\
\bar{\cal H}_p(-+|-+)&=&\bar{\cal H}_p(--|--)  \nonumber\\
&=&-\frac{1}{2}(\epsilon_p^{\prime\prime}+\epsilon_p^{\prime\prime\prime}
+\Gamma_p^{\dag(-+)}+\Gamma_p^{\dag(--)}), \label{eomoo.52}\\
\bar{\cal H}_p(++|-+)&=&\bar{\cal H}_p(+-|--)  \nonumber\\
&=&\frac{1}{2}(\Delta_p^{(+)}+\Delta_p^{(-)}), \label{eomoo.53}\\
\bar{\cal H}_p(-+|++)&=&\bar{\cal H}_p(--|+-)  \nonumber\\
&=&\frac{1}{2}(\Delta_p^{\dag(+)}+\Delta_p^{\dag(-)}), \label{eomoo.54}
\end{eqnarray}
whereas  $\bar{\cal H}_n$ has the block-diagonal structure
\begin{eqnarray}
\bar{\cal H}_n&=&\left(\begin{array}{cc}\underline{\cal H}_n&0\\0&\underline{\cal H}_n
\end{array}\right), \label{eomoo.55}
\end{eqnarray}
in terms of two-by-two matrices, and
\begin{eqnarray}
\underline{\cal H}_n&=&\left(\begin{array}{cc}\epsilon_n^{\prime}+\frac{1}{2}
(\Gamma_n^{(++)}+\Gamma_n^{(+-)})&
\frac{1}{2}(\Delta_n^{(+)}+\Delta_n^{(-)})\\
\frac{1}{2}(\Delta_n^{\dag(+)}+\Delta_n^{\dag(-)})&
-\frac{1}{2}(\epsilon_n^{\prime\prime}+\epsilon_n^{\prime\prime\prime}+
\Gamma_n^{\dag(-+)}+\Gamma_n^{\dag(--)})\end{array}\right). \label{eomoo.56}
\end{eqnarray}
 
We infer from their structure that it is the barred matrices 
that form suitable starting points for the antisymmetrization 
that was the essential step for identifying physical solutions 
for the theory of odd nuclei.  We thus define the
matrices $\bar{\cal H}_{po}$ and $\bar{\cal H}_{no}$,
\begin{eqnarray}
\bar{\cal H}_{po}&=&\frac{1}{2}(\bar{\cal H}_p
-C_p\bar{\cal H}_p\tilde{C}_p), \label{eomoo.57}\\
\bar{\cal H}_{no}&=&\frac{1}{2}(\bar{\cal H}_n
-C_n\bar{\cal H}_n\tilde{C}_n). \label{eomoo.58}
\end{eqnarray} 
Notice that $C_n$ commutes with $\bar{\cal H}_{po}$ and 
$C_p$ commutes with $\bar{\cal H}_{no}$,
but the two Hamiltonians do not commute with each other.
We also have
\begin{eqnarray}
C_n\bar{\cal H}_{no}\tilde{C}_n&=&-\bar{\cal H}_{no},  \label{eomoo.581}\\
C_p\bar{\cal H}_{po}\tilde{C}_p&=&-\bar{\cal H}_{po},  \label{eomoo.582}\\
\end{eqnarray}

To identify the physical basis we introduce an auxiliary Hamiltonian
\begin{equation}
\bar{\cal H}_o(\varepsilon)=\bar{\cal H}_{no}+\varepsilon\bar{\cal H}_{po},
\label{eomoo.583}
\end{equation}
which has the following properties
\begin{eqnarray}
C_p C_n \bar{\cal H}_o(\varepsilon)\tilde{C}_n\tilde{C}_p&=&
-\bar{\cal H}_o(\varepsilon),  \label{eomoo.584}\\
C_p\bar{\cal H}_o(\varepsilon)\tilde{C}_p&=&
\bar{\cal H}_o(-\varepsilon). \label{eomoo.585}
\end{eqnarray}
A quarter of the set of eigenvectors of $\bar{\cal H}_o(1)\equiv
\bar{\cal H}_o$ form the physical basis. From Eq.~(\ref{eomoo.584})
we see that for each positive eigenvalue, $\bar{\cal E}_{Js}
(\varepsilon)>0$ of $\bar{\cal H}_o(\varepsilon)$ which enter
in equation
\begin{equation}
\bar{\cal H}_o(\varepsilon)\Phi_{JM_Js}(\varepsilon)=\bar{\cal E}_{Js}
(\varepsilon)\Phi_{JM_Js}(\varepsilon),   \label{eomoo.586}
\end{equation}
there is a corresponding negative eigenvalue, $-\bar{\cal E}_{Js}
(\varepsilon)$, with the associated eigenvector $C_p C_n\Phi_{JM_Js}
(\varepsilon)$.  Following the standard reasoning of superconductivity 
theory we reject half of the eigenvectors, those belonging to negative
eigenvalues as non-physical.  However that still leaves too many states.

From Eq.~(\ref{eomoo.585}) it follows that the Hamiltonians 
$\bar{\cal H}_o(\varepsilon)$ and $\bar{\cal H}_o(-\varepsilon)$ have 
the same set of eigenvalues, i.\ e., $\bar{\cal E}_{Js}(\varepsilon)
=\bar{\cal E}_{Js}(-\varepsilon)$, with the corresponding eigenvectors
$\Phi_{JM_Js}(-\varepsilon)=C_p\Phi_{JM_Js}(\varepsilon)$.  This has as
a further consequence that each eigenvalue of $\bar{\cal H}_o(0)$,
$\bar{\cal E}_{Js}(0)$ is, apart from the $2J+1$-fold magnetic degeneracy, 
additionally two-fold degenerate with the two eigenvectors 
$\Phi_{JM_Js}(0)$ and $C_p\Phi_{JM_Js}(0)$.  Next we solve 
Eq.~(\ref{eomoo.586}) for $0\leq\varepsilon\leq 1$ and a given $J$.
For $\varepsilon>0$, level $\bar{\cal E}_{Js}(0)$ splits into a pair 
of levels, labeled $\bar{\cal E}_{Js>}(\varepsilon)$ and 
$\bar{\cal E}_{Js<}(\varepsilon)$, distinguishing the larger from the 
smaller value.  We choose the larger of the two eigenvalues as the 
physical one, recognizing its role as the analogue of the sum of 
quasiparticle energies.  We then rely on the no-crossing theorem to 
maintain the ordering of the physical states as we increase the value 
of $\varepsilon$ to unity, that appropriate to the Hamiltonian
$\bar{\cal H}_o(1)\equiv\bar{\cal H}_o$.  Since the number of 
positive eigenvalues of the latter are even, it is a consequence of the 
arguments just given that, for a given $J$, the physical solutions are
the odd-numbered positive eigenvalues, counting from the largest value.

The separation of $\bar{\cal H}_o$ from the original Hamiltonian of 
Eq.~(\ref{eomoo.46}) is achieved by the following decomposition of 
${\cal H}$
\begin{equation}
{\cal H}=\bar{\cal H}_o +\bar{\cal H}_e-\omega+V_{np},\label{eomoo.587}
\end{equation}
where
\begin{eqnarray}
\bar{\cal H}_e&=&\frac{3}{4}({\cal H}_p+{\cal H}_n)+\frac{1}{4}
C_p C_n({\cal H}_p+{\cal H}_n)\tilde{C}_n\tilde{C}_p \nonumber\\
&+&\frac{1}{4}C_p({\cal H}_p -{\cal H}_n)\tilde{C}_p
+\frac{1}{4}C_n({\cal H}_n-{\cal H}_p)\tilde{C}_n. \label{eomoo.588}
\end{eqnarray}
The physical eigenvectors of ${\cal H}$ can further be found using 
again the methods similar to those discussed in Sec.~IIC for odd nuclei.

\subsection{Matrix elements of single particle operators}

Finally, we turn to the problem of deriving a general formula for the transition matrix element, $\langle J'M_{J'} s'|T_{LM_L}|JM_J s\rangle$ of a single-particle operator, $T_{LM_L}$ in a manner analogous to the calculation carried out in  Sec.~\ref{tro}
.  For this purpose, we utilize a formula for the state $|JM_J s\rangle$ (compare Eq.~(\ref{ppk1.statev}))
\begin{eqnarray}
|JM_J s\rangle&=& \frac{1}{\Omega_{(p)}\Omega_{(n)}}\sum_{\sigma\tau pnRM_R r}
\Psi^{(\sigma\tau)}_{JM_J s}(pnRM_R r)a_p^\sigma a_n^\tau |\sigma\tau RM_R r\rangle,   \label{eomoo.59}\\
\Psi^{(\sigma\tau)}_{JM_J s}(pnRM_R r)&=&\langle JM_J s|a_p^\sigma a_n^\tau |
\sigma\tau RM_R r\rangle.  \label{eomoo.60}
\end{eqnarray}
Equations (\ref{eomoo.59}) and (\ref{eomoo.60}) describe a set of orthonormal states, as follows from the equations of motion and the normalization condition
(\ref{eomoo.45}).  We thus derive the formula
\begin{eqnarray}
\langle J'M_{J'}s'|T_{LM_L}|JM_J s\rangle&=&
\frac{1}{\Omega_{(p)}\Omega_{(n)}}\sum_{\sigma\tau pnRM_R rR'M_{R'}r'}
\Psi^{(\sigma\tau)}_{JM_J s}(pnRM_R r) 
\Psi^{(\sigma\tau)}_{J'M_{J'} s'}(pnR'M_{R'} r') \nonumber\\
&&\times\langle\sigma\tau R'M_{R'}r'|T_{LM_L}|\sigma\tau RM_R r\rangle\nonumber\\
&&+\frac{1}{\Omega_{(p)}\Omega_{(n)}}\sum_{\tau pp'nRM_R r}
\Psi^{(-\tau)}_{JM_J s}(pnRM_R r) 
\Psi^{(-\tau)}_{J'M_{J'} s'}(p'nRM_{R} r)t_{p'p} \nonumber\\
&&+\frac{1}{\Omega_{(p)}\Omega_{(n)}}\sum_{\sigma pnn'RM_R r}
\Psi^{(\sigma -)}_{JM_J s}(pnRM_R r) 
\Psi^{(\sigma -)}_{J'M_{J'} s'}(pn'RM_{R} r)t_{n'n} \nonumber\\
&&-\frac{1}{\Omega_{(p)}\Omega_{(n)}}\sum_{\tau pp'nRM_R r}
\Psi^{(+\tau)}_{JM_J s}(pnRM_R r) 
\Psi^{(+\tau)}_{J'M_{J'} s'}(p'nRM_{R} r) \nonumber \\
&&\times t_{\bar{p}\bar{p}'} (-1)^{j_p -m_p +j_{p'}-m_{p'}}  \nonumber\\
&&-\frac{1}{\Omega_{(p)}\Omega_{(n)}}\sum_{\sigma pnn'RM_R r}
\Psi^{(\sigma +)}_{JM_J s}(pnRM_R r) 
\Psi^{(\sigma +)}_{J'M_{J'} s'}(pn'RM_{R} r)\nonumber\\
&&\times t_{\bar{n}\bar{n}'}(-1)^{j_n -m_n +j_{n'}-m_{n'}}. \label{eomoo.61}
\end{eqnarray} 

To apply the Wigner-Eckart theorem requires, in addition to obvious adaptations
of the formulas of Sec.~(\ref{tro}), only the additional formula
\begin{eqnarray}
\Psi^{(\sigma\tau)}_{JM_J s}(pnRM_R r)&=&\sum_l (j_p m_p j_n m_n|lm)
(RM_R lm|JM_J)b_{Js}^{(\sigma\tau)}(pnlRr),   \label{eomoo.62}
\end{eqnarray}
which combines the contents of Eqs.~(\ref{eomoo.6}), (\ref{eomoo.7}) and (\ref{eomoo.60}). We thus find
\begin{eqnarray}
\langle J's'||T_L||Js\rangle&=& 
\frac{1}{\Omega_{(p)}\Omega_{(n)}}\sum_{\sigma\tau lLj_p j_n RrR'r'}
(-1)^{J+R'+l+L}\sqrt{(2J+1)(2J'+1)} \nonumber\\
&&\times\left\{\begin{array}{ccc}J&J'&L\\R'&R&l\end{array}\right\}
\langle\sigma\tau R'r'||T_L||\sigma\tau Rr\rangle \nonumber \\
&&\times b^{(\sigma\tau)}_{Js}(pnlRr)
b^{(\sigma\tau)}_{J's'}(pnlR'r')  \nonumber \\
&&+\frac{1}{\Omega_{(p)}\Omega_{(n)}}\sum_{\tau ll'Lj_p j_{p'}j_n Rr}
(-1)^{R+J'+j_{p'}+j_n}\frac{2j_{p'}+1}{2L+1}\nonumber\\
&&\times\sqrt{(2l+1)(2l'+1)(2J+1)(2J'+1)}
\left\{\begin{array}{ccc}L&J&J'\\R&l'&l\end{array}\right\}
\left\{\begin{array}{ccc}j_p&j_n&l\\l'&L&j_{p'}\end{array}\right\}\nonumber \\
&&\times  t_{p'p}b^{(-\tau)}_{Js}(pnlRr)
b^{(-\tau)}_{J's'}(pnlR'r')  \nonumber \\ 
&&+\frac{1}{\Omega_{(p)}\Omega_{(n)}}\sum_{\tau ll'Lj_p j_{n'}j_n Rr}
(-1)^{R+J'+j_{p}+j_n+l+l'}\frac{2j_{n'}+1}{2L+1}\nonumber \\
&&\times\sqrt{(2l+1)(2l'+1)(2J+1)(2J'+1)}
\left\{\begin{array}{ccc}L&J&J'\\R&l'&l\end{array}\right\}
\left\{\begin{array}{ccc}j_n&j_p&l\\l'&L&j_{n'}\end{array}\right\}\nonumber \\
&&\times  t_{n'n}b^{(\sigma -)}_{Js}(pnlRr)
b^{(\sigma -)}_{J's'}(pn'lR'r')  \nonumber \\ 
&&+\frac{1}{\Omega_{(p)}\Omega_{(n)}}\sum_{\tau ll'Lj_p j_{p'}j_n Rr}
(-1)^{R+J'+j_{p'}+j_n}\frac{2j_{p'}+1}{2L+1}\nonumber\\
&&\times\sqrt{(2l+1)(2l'+1)(2J+1)(2J'+1)}
\left\{\begin{array}{ccc}L&J&J'\\R&l'&l\end{array}\right\}
\left\{\begin{array}{ccc}j_p&j_n&l\\l'&L&j_{p'}\end{array}\right\}\nonumber \\
&&\times  (-1)^{j_p+j_{p'}-L}t_{pp'}b^{(+\tau)}_{Js}(pnlRr)
b^{(+\tau)}_{J's'}(pnlR'r')  \nonumber \\ 
&&+\frac{1}{\Omega_{(p)}\Omega_{(n)}}\sum_{\tau ll'Lj_p j_{n'}j_n Rr}
(-1)^{R+J'+j_{p}+j_n+l+l'}\frac{2j_{n'}+1}{2L+1}\nonumber\\
&&\times\sqrt{(2l+1)(2l'+1)(2J+1)(2J'+1)}
\left\{\begin{array}{ccc}L&J&J'\\R&l'&l\end{array}\right\}
\left\{\begin{array}{ccc}j_n&j_p&l\\l'&L&j_{n'}\end{array}\right\}\nonumber \\
&&\times  (-1)^{j_n+j_{n'}-L}t_{nn'}b^{(\sigma -)}_{Js}(pnlRr)
b^{(\sigma -)}_{J's'}(pn'lR'r').  \label{eomoo.63}
\end{eqnarray}
Once again we have a clear separation into collective and single-particle contributions.

\section{Summary and conclusions}

A linearized version of the equations of motion approach to the nuclear
many-body problem, considered as a generalization of traditional core-particle coupling models has proved its worth in a number of recent applications to deformed odd nuclei.  In this method, the basic object studied is a single-particle coefficient of fractional parentage (CFP) relating the states of the even nuclei to those of a neighboring odd nucleus.  

In this paper we showed how the 
same general method can be applied to odd-odd nuclei.  We started with a
review of the formalism for odd nuclei, since it plays an essential
role in some of the considerations that follow.  We then showed that there are three possible formulations for the odd-odd case, two of which we label as sequential and a third as symmetrical, terms that characterize the way in which we couple an extra neutron (or neutron hole) and an extra proton (or proton hole) to nearby even nuclei, treated as cores.  First we study in detail the case where we initially  
couple the odd neutron to the even cores, an example of our method for
odd nuclei.  We then couple the odd proton to the odd neutron nuclei,
introducing new CFP for this relationship, and 
making essential use of the odd neutron calculations for energies and 
CFP.  The second
sequential method, not discussed in detail, reverses the order of the
odd-particle couplings.  In the symmetrical coupling, we first couple
the two odd particles together and study directly the relationship of the odd-odd nucleus to the core even nuclei by means of two-particle
CFP.  In principle all three methods are equivalent, but in practice 
results will differ owing to the need to approximate.  In this regard,
the existence of alternatives that may be compared may be of some practical advantage.

Because of the presence of pairing interactions the equations for the
odd-odd case yield four times as many solutions as are physical.  In the sequential method, the problem of choosing physical solutions can be solved by sequential use of essentially the same method as for the odd case.  For the symmetrical coupling case, a more elaborate method
has been devised.

Concerning applications, approximate versions of the sequential method
have already been carried out \cite{STAR,Kioke}.  The symmetrical approach remains to be tried.

\acknowledgments
One of the authors (S.G.R) would like to acknowledge partial support by the Polish Committee for Scientific Research(KBN) under  Contract No. P03B 014 21.  Two of us (A.K) and (S.G.R) are grateful to the Institute for Nuclear Theory at the University of Washington (Seattle)
where this work was initiated while we were both in residence.

\end{document}